\documentclass[journal=jpccck,manuscript=article]{achemso}
\usepackage[english]{babel}
\usepackage{amsmath,amsfonts,amssymb}
\usepackage{booktabs}
\usepackage{threeparttable}
\usepackage{dcolumn}
\usepackage{units}
\usepackage{braket}
\usepackage{enumerate}
\usepackage{subfig}

\newcolumntype{.}{D{.}{.}{0}}
\newcolumntype{;}{D{.}{.}{-1}}

\author{Gonzalo Angulo}
\email{gangulo@ichf.edu.pl}
\affiliation{Institute of Physical Chemistry, Polish Academy of Sciences, 44/52 Kasprzaka, 01-224 Warsaw, Poland}

\author{Marta Brucka}
\affiliation{Department of Organic Chemistry, University of Geneva, 30, Quai Ernest Ansermet, CH-1211 Geneva, Switzerland}

\author{Mario Gerecke}
\affiliation{Department of Chemistry, Humboldt Universit\"{a}t zu Berlin, D-12489 Berlin, Germany}

\author{G\"unter Grampp}
\affiliation{Institute of Physical and Theoretical Chemistry, Graz University of Technology, Stremayrgasse 9, 8010 Graz, Austria}

\author{Damien Jeannerat}
\affiliation{Department of Organic Chemistry, University of Geneva, 30, Quai Ernest Ansermet, CH-1211 Geneva, Switzerland}

\author{Jadwiga Milkiewicz}
\affiliation{Institute of Physical Chemistry, Polish Academy of Sciences, 44/52 Kasprzaka, 01-224 Warsaw, Poland}

\author{Yavor Mitrev}
\affiliation{Department of Organic Chemistry, University of Geneva, 30, Quai Ernest Ansermet, CH-1211 Geneva, Switzerland}
\altaffiliation{Institute of Organic Chemistry with Centre of Phytochemistry, Bulgarian Academy of Sciences, Acad.\ G.\ Bonchev Str., bl 9, 1113 Sofia, Bulgaria}

\author{Czes\l aw Radzewicz}
\affiliation{Institute of Physical Chemistry, Polish Academy of Sciences, 44/52 Kasprzaka, 01-224 Warsaw, Poland}
\altaffiliation{Institute of Experimental Physics, Faculty of Physics, University of Warsaw, ul.\ Pasteura 5, 02-093 Warsaw, Poland}

\author{Arnulf Rosspeintner}
\email{arnulf.rosspeintner@unige.ch}
\affiliation{Department of Physical Chemistry, University of Geneva, 30, Quai Ernest Ansermet, CH-1211 Geneva, Switzerland}

\author{Eric Vauthey}
\affiliation{Department of Physical Chemistry, University of Geneva, 30, Quai Ernest Ansermet, CH-1211 Geneva, Switzerland}

\author{Pawe\l\ Wnuk}
\affiliation{Institute of Physical Chemistry, Polish Academy of Sciences, 44/52 Kasprzaka, 01-224 Warsaw, Poland}
\altaffiliation{Institute of Experimental Physics, Faculty of Physics, University of Warsaw, ul.\ Pasteura 5, 02-093 Warsaw, Poland}

\title[]{Characterization of Dimethylsulfoxide / Glycerol Mixtures: A Binary Solvent System for the Study of ``Friction-Dependent" Chemical Reactivity}

\begin{document}
\begin{abstract}
\noindent The properties of binary mixtures of dimethylsulfoxide and glycerol, measured by several techniques, are reported. Special attention is given to those properties contributing or affecting chemical reactions. In this respect the investigated mixture behaves as a relatively simple solvent and it is especially well suited for studies on the influence of viscosity in chemical reactivity. This is due to the relative invariance of the dielectric properties of the mixture. However, special caution must be taken with specific solvation, as the hydrogen-bonding properties of the solvent changes with the molar fraction of glycerol.
\end{abstract}

\section{Introduction}
Most of the theories describing chemical reactions and dynamics in liquid solutions are expressed in terms of macroscopic properties of the system.\cite{nitzan_2006, tomasi_CR_2005} This is because they often derive from or are related to continuous descriptions of liquids. Reaction-diffusion problems,\cite{burshtein_ACP_2004} solvation dynamics,\cite{horng_JPC_1995, sajadi_JPCA_2009} vibrational relaxation of highly excited molecular states,\cite{pigliucci_JPCA_2007, braem_PCCP_2012} isomerization reactions,\cite{weigel_PCCP_2012, ruetzel_PNAS_2014} electron transfer,\cite{barzykin_ACP_2002, matyushov_JCP_2004, rosspeintner_JACS_2014} proton transfer,\cite{park_JCP_2009} or energy transfer,\cite{grampp_PPS_2009} are thus correlated to solvent parameters like viscosity, dielectric constant, density or refractive index. Therefore, it is often desirable to vary one of them keeping the others constant fulfilling the \emph{ceteris paribus} condition.\cite{scerri_2015} Simultaneously, it is also necessary to know if the system under study, the probe molecule or the reactants, are prone to build H-bonds with the solvent or are inert to them. The difficulty arises because most of the homologous series of solvents change several of their properties on increasing the chain length.\cite{marcus_1998} For example, in linear alcohols or nitriles the dielectric constant decreases at the same time that the viscosity increases. Moreover, in the study of diffusion influenced reactions, the key control parameter is the viscosity and there are no solvent families that offer a wider range of this property than alcohols. If the reaction of interest is electron transfer, the Marcus solvent reorganization energy is proportional to the Pekar factor, a function of the refractive index and the dielectric constant;\cite{marcus_JCP_1956} thus changing viscosity to determine the influence of diffusion in electron transfer reactions requires keeping this factor constant. The use of alcohols having the advantage of providing with a wide range of viscosities introduces changes in the other parameters. Another possibility is to change either temperature or pressure,\cite{ilichev_BBPC_1997} but all the solvent properties, as well as the intrinsic reaction rate, are sensitive in larger or lesser degree to them. The only remaining solution in this case is to use solvent mixtures.\cite{gardecki_CPL_1999} A wise selection of the binary components may provide with the control of one of the four important parameters mentioned above. However, there are also problems associated with those mixtures. First, if the dielectric constants of the components are too disparate, the solvation shell around the solutes can be enriched with one of them.\cite{petrov_HEC_2006, suppan_JCSFT1_1987} Second, if the H-bonding properties of the components are different and the solute is sensitive to them, the solvation properties may change dramatically too. Finally, micro-environments may show up in some of these mixtures that can significantly affect the diffusion of the reactants.\cite{roux_JSC_1980, lara_JSC_1981}\\
One of the most successfully used mixtures to study diffusion-assisted reactions consists of dimethylsulfoxide (DMSO) and glycerol (GLY). It has been employed in the study of fluorescence quenching reactions by electron transfer, as well as in the recombination of the products.\cite{neufeld_JCP_2002, angulo_JPCA_2003, gladkikh_PCCP_2003, gladkikh_JPCA_2007, rosspeintner_CEJ_2007, angulo_CEJ_2010, koch_JACS_2012, rosspeintner_JACS_2012} It has served to test and articulate the encounter theories that have demonstrated to be very accurate in the quantitative description of bimolecular reactions. Recently, it was employed in the comparison with room temperature ionic liquids to clarify the role of diffusion, leading to the conclusion that nothing peculiar is to be found in the latter case, at least with respect to photo-induced bimolecular electron transfer.\cite{koch_JACS_2012} In regard to dielectric constants, refractive index, densities and molecular sizes, the two solvents are very similar (cf.\ Table \ref{tab:pure_props}). Only viscosity differs greatly, as well as the H-bonding donation ability, in terms of the Kamlet-Taft (KT) empirical parameters ($\alpha$, $\beta$, $\pi^*$).\cite{kamlet_JOC_1983} It is the aim of this communication, not only to summarize these macroscopic properties, but also to shine light on the solvation properties at the molecular scale.\\
Two research groups have recently reported on the solvation properties of DMSO-GLY finding the presence of micro-heterogeneities.\cite{chattoraj_JCP_2013, kaur_JPCB_2014, koley_PCCP_2014} In ref.~\citenum{chattoraj_JCP_2013} confocal fluorescence microscopy was used. From the fluorescence correlation curves (FCS) and the burst integrated fluorescence lifetimes (BIFL) distributions, it was concluded that there are enriched environments of DMSO through which the fluorescent coumarin probes diffuse faster than would correspond to the macroscopic viscosity. In references \citenum{kaur_JPCB_2014} and \citenum{koley_PCCP_2014} the steady-state fluorescence spectra and the time-resolved fluorescence decays obtained by means of nanosecond resolved fluorescence of several coumarins were reported. From the analysis of the solvation dynamics and of the anisotropy decays a similar conclusion as in reference \citenum{chattoraj_JCP_2013} was reached.\\ 
If micro-environments are present in these mixtures, as suggested by these works,\cite{chattoraj_JCP_2013, kaur_JPCB_2014, koley_PCCP_2014} they are expected to be relevant to molecular diffusion, either translational or rotational, which should manifest in distinct fluorescence anisotropy or time resolved diffusion dynamics. On the other side, none of the previous studies dealing with bimolecular reactions in these mixtures has revealed such a peculiar behavior. The results of the measurements of both, bulk and microscopic, properties of this mixture, obtained from multiple experimental techniques, do not require invoking the presence of micro-environments, in order to be rationalized. We will discuss and compare the previous reports in view of the current findings. 

\section{Materials and Methods}

\begin{table*}
\centering
\begin{threeparttable}[b]
		\caption{Solvent properties of the pure solvents at 20$^\circ$C.}
				\label{tab:pure_props}
		\begin{tabular}{lcccccccccccc}
			\toprule
		    & $\mu$ & $V$ 		& $S$ 		& $\rho$	& $\eta$	& $\epsilon$	& $n_{\rm D}$	& $f(\epsilon)$	&$f(n_{\rm D}^2)$ & $\alpha$ & $\beta$ & $\pi^*$  \\ 
		    &  (D)      &  (\AA$^3$) 	& (\AA$^2$) 	& (\unitfrac[]{kg}{L})	& (cP)	&  			&			& & & & &  \\
		    		\midrule
	DMSO & 4.06   & 69.4    & 102.0     & 1.26 & 2.20 & 46.02 & 1.4793 & 0.484 & 0.121 & 0.00     & 0.76 & 1.00    \\ 
	GLY     & 4.21   & 85.2    & 127.1   & 1.10 & 1412 & 42.50 & 1.4746 & 0.483 & 0.120 & 1.21 & 0.51 & 1.00\\
		\bottomrule      
		\end{tabular}
		\begin{tablenotes}
		\item [] $\rho$, $\eta$, $\epsilon$ and $n_{\rm D}$ are the density, dynamic viscosity, dielectric constant and refractive index of the solvent at 20$^\circ$C (the $\eta$ value for glycerol was taken from ref.~\citenum{riddick_1986}). $\mu$, $V$ and $S$ are the dipole moment, van der Waals volume and surface, respectively (see ref.~\citenum{marcus_1998}). $f(x)$ is the relevant contribution from the dielectric constant, $\epsilon$ and refractive index, $n_{\rm D}$, to the Onsager reaction field, and is defined as $f(x) = (x-1)/(2x+1)$.\cite{onsager_JACS_1936} $\alpha$, $\beta$ and $\pi^*$ are the solvatochromic parameters for specific and unspecific solute-solvent interactions according to the Kamlet-Taft scale (see ref.~\citenum{marcus_JSC_1991}; note that we have recalculated the $\pi^*$ value of glycerol using the equation given in this reference using the dipole moment given in reference \citenum{marcus_1998}).
		\end{tablenotes}

\end{threeparttable}
\end{table*}

\subsection{Materials}
Dimethylsulfoxide (Alfa Aesar, 99.9+\%) and glycerol (Alfa Aesar, ultrapure HPLC grade) were used as received and stored water-free under argon. DMSO-d6 (Cambridge Isotope Labs, D, 99.9\%) was used as received. Coumarin 151 (Exciton), 152A (Exciton), 153 (Radiant Dyes) and 500 (Exciton) were used as received and Auramine O was recrystallized twice from ethanol. A sketch of the chemical structures can be found in the supporting information.

\subsection{Experimental Methods}
Kinematic viscosities, $\nu$, of DMSO and of the binary mixtures were measured at 20.0$\pm0.1^\circ$C using an Ubbelohde viscometer (0C, IC and IIC viscometers from Schott Ger\"{a}te). The densities were determined at the same temperature using \unit[50]{mL} pycnometers with \unit[0.1]{mL} error.
Dielectric constants, $\epsilon$, of the pure solvents and binary mixtures were measured at 20.0$\pm0.2^\circ$C using a home-built LC circuit operating at approximately \unit[400]{kHz}. The apparatus was calibrated using a series of organic solvents and the calibration curve deviates only slightly from the Thompson equation for the ideal circuit.\cite{bonilla_JCE_1977} Refractive indices, $n_{\rm D}^{20}$, were measured at 20$\pm0.2^\circ$C using an Abbe refractometer from Atago (model 1T).\\
Absorption spectra were recorded on a Cary 50 spectrometer at room temperature (22$\pm2^\circ$C), while steady-state emission spectra were recorded on a FluoroMax-4 (Jobin Yvon) at 20$\pm0.2^\circ$C and corrected using a set of secondary emissive standards.\cite{gardecki_AS_1998}\\
Time-resolved fluorescence experiments were performed on five different set-ups. Ultrafast time-resolved anisotropy decays were measured on a single-wavelength fluorescence up-conversion set-up with a full width of half maximum of the instrument response function (IRF) of \unit[200]{fs}, described in more detail in reference \citenum{morandeira_JPCA_2004}. In order to obtain reliable information on the solvation dynamics, and given the vast discrepancies encountered in this type of experiments (see e.~g.\ reference \citenum{zhang_RSI_2011} for a comparison of previous literature data) we opted for performing 
time-resolved Stokes shift measurements on three time-resolved broadband fluorescence up-conversion set-ups (FLUPS) two of which (Ernsting group in Berlin and Vauthey group in Geneva) based on the set-up described in references\ \citenum{zhao_PCCP_2005, zhang_RSI_2011, sajadi_APL_2013}, and one (Laser Center of the PAS in Warsaw) described in references\ \citenum{bialkowski_JPPA_2012, wnuk_PCCP_2014}, with a time resolution of \unit[100-170]{fs}. It is worth noting, that the dynamic Stokes shifts obtained on all three broadband set-ups are in excellent agreement among each other (cf.\ SI for a comparison of both, spectra and peak-shifts).\\
 For dynamics longer than \unit[1]{ns} single wavelength measurements on a home-built time-correlated single photon counting (TCSPC) apparatus (time-resolution of \unit[200]{ps}), described in reference \citenum{muller_CPL_2000}, were used.\\
All NMR samples were recorded at \unit[298]{K} on a Bruker \unit[500]{MHz} $^1$H NMR Larmor frequency spectrometer equipped with a DCH helium-cooled detection probe equipped with a z-gradient coil with a maximum nominal gradient strength of \unitfrac[65]{G}{cm}. The samples were prepared in standard \unit[5]{mm} NMR tubes, with the sample height being below \unit[4]{cm} in order to avoid convection caused by nonhomogeneous heating of the NMR tube above the level corresponding to the top of the probehead. The NMR self-diffusion measurements where obtained using a longitudinal encode-decode experiment including a double stimulated echo\cite{jerschow_JMR_1996, jerschow_JMR_1997} and with the Bruker pulse program \emph{dstegp3}. The diffusion delay $\Delta$ was set to \unit[100]{ms}, the diffusion-encoding field-gradient pulse $\delta$ was set to \unit[6]{ms}; with spoiling gradient of \unit[1]{ms}. The top amplitude of the sine-bell shaped diffusion gradient pulses, $G_i$, ranged from 1.3 to \unitfrac[55.3]{G}{cm} in 16 linear steps. The spectral window was set to \unit[15]{ppm} and the free induction decay was acquired with \unit[32]{k} points. The recovery delay was set to \unit[18]{s} to ensure complete relaxation. Each spectrum was obtained by summation of 8 transients recorded after 16 stabilization acquisitions (Òdummy scansÓ) resulting in a total experimental time of \unit[49]{min}. The FIDs were Fourier-transformed after multiplication with an exponential function with a coefficient of \unit[5]{Hz}.

\subsection{Analysis Methods}
The KT parameters of the solvent mixtures were obtained by analyzing the absorption spectra of 4 coumarins with known KT parameters (cf.\ Table \ref{tab:kamlet}) by performing a multilinear regression analysis for each solvent mixture.
\begin{align}
\nu_{ij}^{\rm exp} & = \nu_{i{\rm DMSO}}^{\rm exp} + a_i \Delta \alpha_j + b_i \Delta \beta_j + s_i \Delta \pi_j^* \label{eq:KAT}\\
\Delta x_j & = x_j - x_{\rm DMSO} \  \text{with}\ x \in \{ \alpha, \beta, \pi^* \}
\end{align}
$\nu_{ij}^{\rm exp}$ is the wavenumber of the maximum of the absorption spectrum of solute $i$ in solvent $j$ in the transition dipole moment representation,\cite{angulo_SAPA_2006} obtained by fitting a \emph{lognorm} function to the experimental data. In order to keep the experimental scatter as low as possible, $\nu_{ij}^{\rm exp}$ as a function of the mole fraction of glycerol, $x_{\rm GLY}$, was fitted with 2$^{\rm nd}$-order polynomials, which were further used as inputs in eq.~\eqref{eq:KAT} (cf.\ SI for the polynomial fits).

\begin{table}
\centering
\caption{Solute parameters for 4 coumarins obtained from Kamlet-Taft analysis of their low energy absorption maxima.}
		\label{tab:kamlet}
	\begin{threeparttable}[b]
	\begin{tabular}{lcc.c}
		\toprule
		      & $\nu_0$ (kK)\tnote{1} & $a$ (kK) & \multicolumn{1}{c}{$b$ (kK)} & $s$ (kK) \\ 
		\midrule
		C151\tnote{2}    & 28.63    & -0.05  & -2.36  & -0.84  \\ 
		C152A\tnote{2}   & 25.93    & -0.35  & -0.23  & -1.03\\
		C153\tnote{3}    & 24.90    & -0.54  & 0.15   & -1.78  \\ 
		C500\tnote{2}    & 27.27    & -0.32  & -1.33  & -0.93  \\ 
		\bottomrule
	\end{tabular}
	\begin{tablenotes}
		\item [1] \unit[1]{kK} = \unit[1000]{cm$^{-1}$}
		\item [2] From reference \cite{das_JPCA_2006}
		\item [3] From reference \cite{molotsky_JPCA_2003}
		\end{tablenotes}
\end{threeparttable}

\end{table}

The ultrafast rotational dynamics of C153 were recorded with the single wavelength up-conversion set-up between \unit[520-540]{nm} by changing the polarization of the pump beam using a halfwave-plate. Parallel, $I_{||}(t)$, and perpendicular signals, $I_{\perp}(t)$, were collected in successive measurements and had a maximal intensity of approx.\ \unit[20-30'000]{cts} (cf.\ SI for exemplary time-traces). Rotational dynamics of C153 in the nanosecond range were recorded using TCSPC by exciting with parallel polarized light and measuring successively $I_{||}(t)$ and $I_{\perp}(t)$ (up to \unit[30'000]{cts} in the maximum) by turning the analyzer in the emission path. Since all rotational times were significantly longer than the instrument response functions (IRF), the anisotropy dynamics, $r(t)$, was calculated and fitted with biexponential decays without accounting for convolution with the IRF.
\begin{equation}
r(t) = \dfrac{I_{||}(t) - I_{\perp}(t)}{I_{||}(t) +2 I_{\perp}(t)}.
\end{equation}
The quality of the fits was judged from the resulting reduced $\chi_{\rm r}^2$.\cite{wahl_BC_1979}\\
The dynamic solvent response was analyzed by monitoring the peak position, $\nu_{\rm p}(t)$, of a \emph{lognorm} fit (see ref.~\citenum{horng_JPC_1995} for the definition of the parameters) to the time-dependent spectra at each time-step (cf.\ SI for exemplary broadband spectra and frequency shifts on the 3 used set-ups), as follows
\begin{equation}
\nu_{\rm p}(t|t>0.5{\rm ps}) = \nu(\infty) + \sum_i \Delta \nu_i \exp\left(-t/\tau_i \right).
\label{eq:St}
\end{equation}
The translational diffusion coefficients from pulsed field gradient (PFG) NMR experiments were obtained by fitting
\begin{equation}
I = I(0) \exp \{ -D \left(2\pi\gamma \delta G_i \right)^2 \left(\Delta - \frac{1}{3}\delta \right) 10^4 \},
\end{equation}
to the signal intensities, $I$, obtaining the initial intensity, $I(0)$, and the diffusion coefficient, $D$ in \unitfrac[]{m$^2$}{s}, and using \unitfrac[$\gamma = 4.258\cdot 10^3$]{Hz}{G}.

\section{Results and Discussion}

\subsection{Macroscopic Solvent Parameters}
As is common for mixtures of organic solvents none of the macroscopic properties here under study show a linear dependence with the molar fraction. The dependence of the density, refractive index and dielectric constant of the solvent mixtures on the mole fraction of glycerol, $x_{\rm GLY}$ was evaluated using a Kister-Redlich type equation (eq.\ \eqref{eq:kister}),\cite{redlich_IEC_1948} while the solvent viscosity was described using the equation of Nissan-Grunberg (eq.\ \eqref{eq:nissan}):\cite{grunberg_N_1949}
\begin{equation}
y_{\rm m}  =  \sum\limits_{i=1}^2 y_i x_i + x_1x_2 \sum\limits_{j=1}^n a_j (2x_1-1)^{j-1}
\label{eq:kister}
\end{equation}
\begin{equation}
\ln(y_{\rm m})  =  \sum\limits_{i=1}^2 x_i \ln(y_i) + x_1x_2 a_1
\label{eq:nissan}
\end{equation}
Here $y_{\rm m}$ denotes the property of interest of the solvent mixture, $x_i$ is the molar fraction of component $i$ ($i=1$ is used for glycerol in eq.\ \eqref{eq:kister}) and $a_i$ are coefficients given in Table \ref{tab:fit_pars}.

\begin{table}[!htp]
	\begin{center}
		\caption{Fitting parameters for the macroscopic solvent properties of the solvent mixture at 20$^\circ$C.}
				\label{tab:fit_pars}
	\begin{threeparttable}[b]
	\begin{tabular}{ccccc}
			\toprule
		        & $y_1$ 	& $y_2$ 	& $a_1$  & $a_2$      \\
		        \midrule
		        & \multicolumn{4}{c}{Kister-Redlich eq.~\eqref{eq:kister}}     \\ 
 
	$\rho$     	& 1.260          & 1.100  & 0.0235 &    0     \\ 
	$n_{\rm D}^{20}$& 1.4746          & 1.4794  & 0.0084   & -0.0015 \\ 
	$\epsilon$& 42.4         & 45.9 & 26.4   & -22.4 \\ [1ex]
		        & \multicolumn{4}{c}{Nissan-Grunberg eq.~\eqref{eq:nissan}}        \\ 
	$\eta$     & 1420            & 2.20     & -0.691 &        \\
		\bottomrule      
		\end{tabular}
		\end{threeparttable}
	\end{center}
\end{table}

\begin{figure}
\centering
	\includegraphics[scale=1.5]{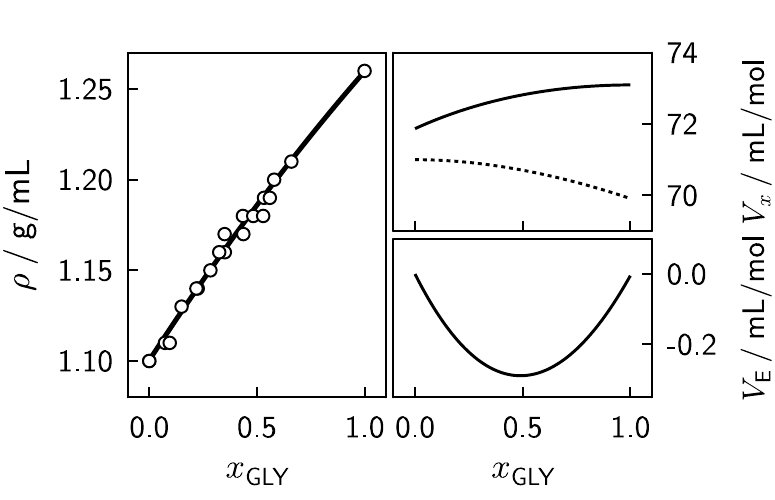}
	\caption{Mass density, $\rho$, excess volume, $V_{\rm E}$, and partial molar volumes, $V_x$, (full line - GLY, dashed line - DMSO) of the binary mixture of DMSO/GLY at 20$^\circ$C as a function of the mole fraction of GLY. }
\label{fig:density}
\end{figure}
\subsubsection{Density}
The densities of the binary mixtures allow to extract the excess volume of the mixture (cf.\ Figure~\ref{fig:density}). It shows a minimum at $x_{\rm GLY} = 0.48$ of about $-0.4\%$ of the molar volume of the mixture at this molar fraction (\unitfrac[71.7]{mL}{mol}), thus departing only slightly from the behavior of an ideal mixture. The partial molar volumes of both components exhibit no extremes at intermediate values of the molar fraction, and change monotonously with increasing the glycerol content: decreasing for DMSO and increasing for GLY. Pronounced extremes in mixtures of alcohols and amines with water in this quantity have been attributed to the presence of microphase segregation.\cite{roux_JSC_1980, lara_JSC_1981, shirota_JCP_2000} It is possible that in our measurements this effect is absent due to the strong basicity of DMSO, as stated by El Seoud in his analysis of solvation in DMSO/water mixtures.\cite{elseoud_PAC_2007}

\subsubsection{Refractive Index \& Dielectric Constant}
\begin{figure}[htp]
\centering
	\includegraphics[scale=1.5]{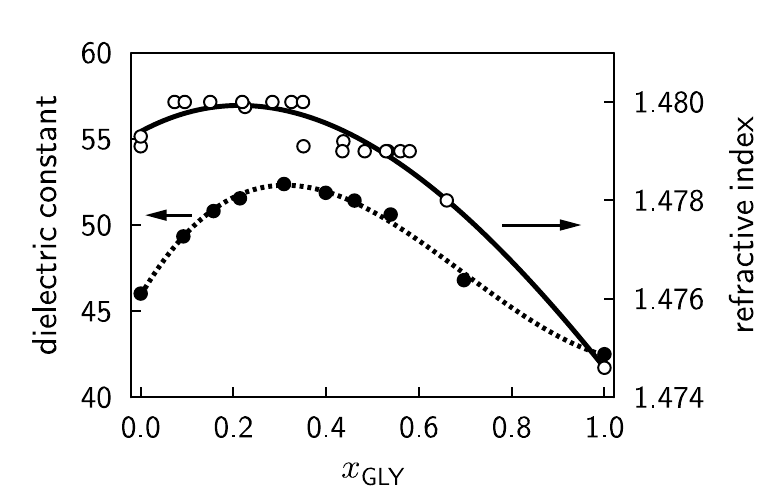}
	\caption{Refractive index ($\circ$) and dielectric constant ($\bullet$) of the DMSO/GLY mixture at 20$^\circ$C as a function of the mole fraction of GLY.}
\label{fig:refractive}
\end{figure}

The refractive indices of the mixtures vary smoothly by less than 0.006 showing a maximum at $x_{\rm GLY}\approx 0.2$ (cf.\ Fig.~\ref{fig:refractive}). On the other hand, the dependence of the static dielectric constant with the molar fraction of glycerol shows a clear maximum with a value of 52 at roughly $x_{\rm GLY} = 0.3$. This maximum is 6 units larger than the value of the more polar of the two constituents, i.~e.\ DMSO (cf.\ Fig.~\ref{fig:refractive}). This kind of deviation from the linear behavior in the dielectric constant is usually attributed to an increase in the number of dipoles in the liquid with respect to the pure solvents. For reasons, hitherto unknown, our measurements are at odds with those presented in ref.~\citenum{jie_JPCA_2013}, where no excess dielectric constant could be observed. A maximum of the static dielectric constant can be interpreted as appearing at the mixture composition at which the number of highly polar heterodimers in the solvent is largest.\cite{jie_JPCA_2013} However, as pointed out by Kaatze et al.\cite{kaatze_IJT_2014} these data are in principle not enough to extract this kind of conclusion as the orientational correlation between dipoles, composed of at least three contributions in binary mixtures, may also change with solvent composition. It is in any case to be considered that the hydrogen-bonding of GLY is strongly perturbed with increasing DMSO content. This is especially important in the case of a solvent such as DMSO, which exhibits a high H-bonding accepting ability. However, we emphasize that from the point of view of continuum theories of the medium, such as the Born model, and thus the study of chemical reactions this mixture represents an almost invariant combination for the reorganization energy, $\lambda_{\rm solv} \propto 1/n_{\rm D}^2 - 1/\epsilon$ of the solvent and the free enthalpy of electron transfer reactions, $\Delta G_{\rm et} \propto 1/\epsilon$. The change with the GLY content is minimal for both quantities.

\subsubsection{Viscosity}
\begin{figure}[htp]
\centering
	\includegraphics[scale=1.5]{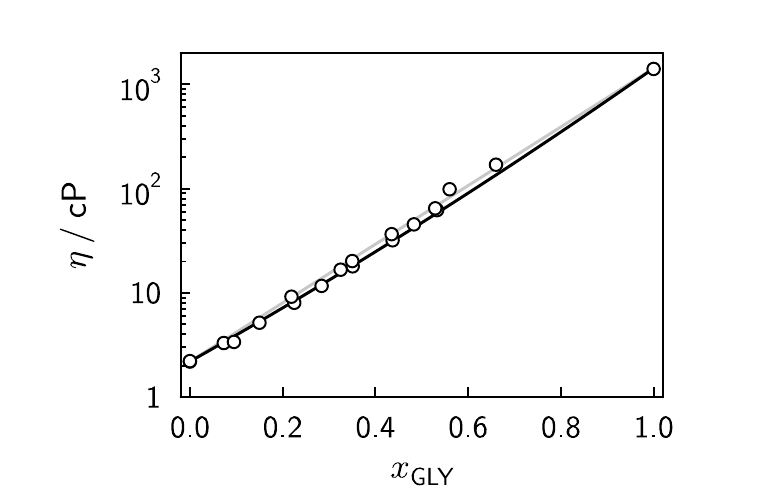}
	\caption{Viscosity of the binary DMSO/GLY mixture at 20$^\circ$C as a function of the mole fraction of GLY. The black line shows the fit of the Nissan-Grunberg equation (eq.~\eqref{eq:nissan}) with the parameters from Table~\ref{tab:fit_pars}, while the grey line shows ideal behaviour.}
\label{fig:viscosity}
\end{figure}

As can be seen in Fig.~\ref{fig:viscosity} the viscosity is the sole quantity of those reported so far that changes greatly with solvent composition. However, no extrema are found in the entire range of molar fractions studied, as the change is monotonous. In combination with the quasi-invariance of the dielectric properties of the mixture, DMSO/GLY mixtures seem at this point an ideal system for studies requiring only variations of the viscosity. It is obvious, however, that some other properties change as the H-bonding properties of the components are significantly different (see Table~\ref{tab:pure_props}). 

\subsubsection{Kamlet-Taft Parameters \label{sec:KAT}}

\begin{figure}[!htp]
\centering
	\includegraphics[scale=1.5]{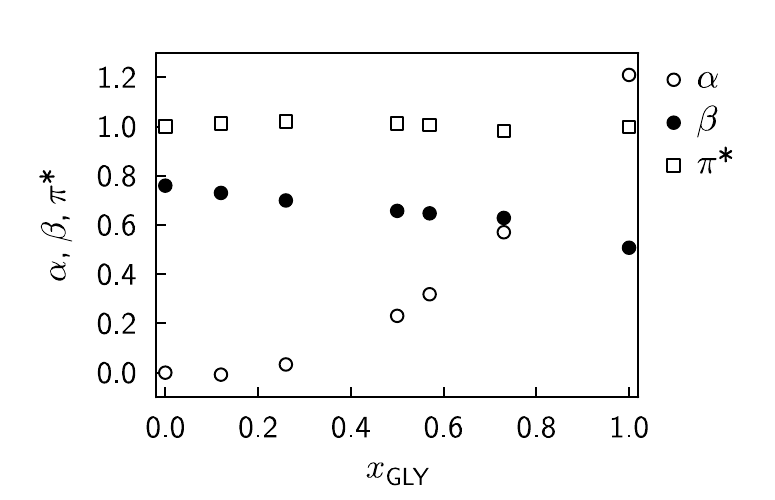}
	\caption{Dependence of the three Kamlet-Taft parameters on the mole fraction of GLY in the DMSO/GLY mixture. The values at $x_{\rm GLY}=1$ were taken from Table~\ref{tab:pure_props}.}
\label{fig:KAT}
\end{figure}

The Kamlet-Taft parameters belong to an empirical scale based on the assumption of a linear energy response of different solvation processes.\cite{kamlet_JOC_1983} It splits the response of an observable, like the position of the electronic absorption band of a fluorophore, to the process of solvation into three major contributions from interactions with the solvent: the ability of the solvent to donate H in H-bonds with the solute, $\alpha$, to accept them, $\beta$, and the polarity/polarizability of the solvent, $\pi^*$. Thus, it allows for the study of solvation when specific solvation effects due to H-bonding are important, and reveals through solute-dependent sensitivity coefficient, $a$, $b$ and $s$, which portion of the relaxation is due to each of the processes in absolute energy terms. This is not the sole existing scale of this kind, as for example the one due to Catal\'{a}n and co-workers is also extensively and successfully used throughout the literature.\cite{catalan_JPCB_2009}\\
The absorption spectra of the coumarins used to obtain the KT parameters of the mixture are shown in the SI. The shape of the spectra does not change with the glycerol content, which provides consistency to the analysis method used. We have chosen to extract the KT parameters solely from the absorption spectra since it is not guaranteed that at the largest viscosities used here the stationary emission spectra completely originate from the relaxed excited state. This is especially the case for the two coumarins with short lifetimes, i.~e.\ C152 and C152A.\\
The variation of the three KT parameters is plotted in Fig.~\ref{fig:KAT}. As expected, the largest change is observed for $\alpha$ while the smallest change is observed for $\pi^*$, which can be related to the relatively small change in dielectric constant as described above. In this case again a very mild maximum is observed around $x_{\rm GLY} = 0.3$ although much less pronounced than for the dielectric constant. This is perfectly congruent as the changes in $\pi^*$ monitor rather the Onsager function, $f(\epsilon) - f(n_{\rm D}^2)$, where $f(x) = (x-1)/(2x+1)$, than directly the dielectric constant. Therefore, both the macroscopic dielectric constant measurements and the molecular probe, via the KT parameters, report similar trends. $\beta$, drops monotonically upon going from DMSO to GLY.

\subsection{Dynamic Properties}

\subsubsection{Solvation Dynamics}
\begin{figure}[h!]
\centering
	\includegraphics[scale=1.5]{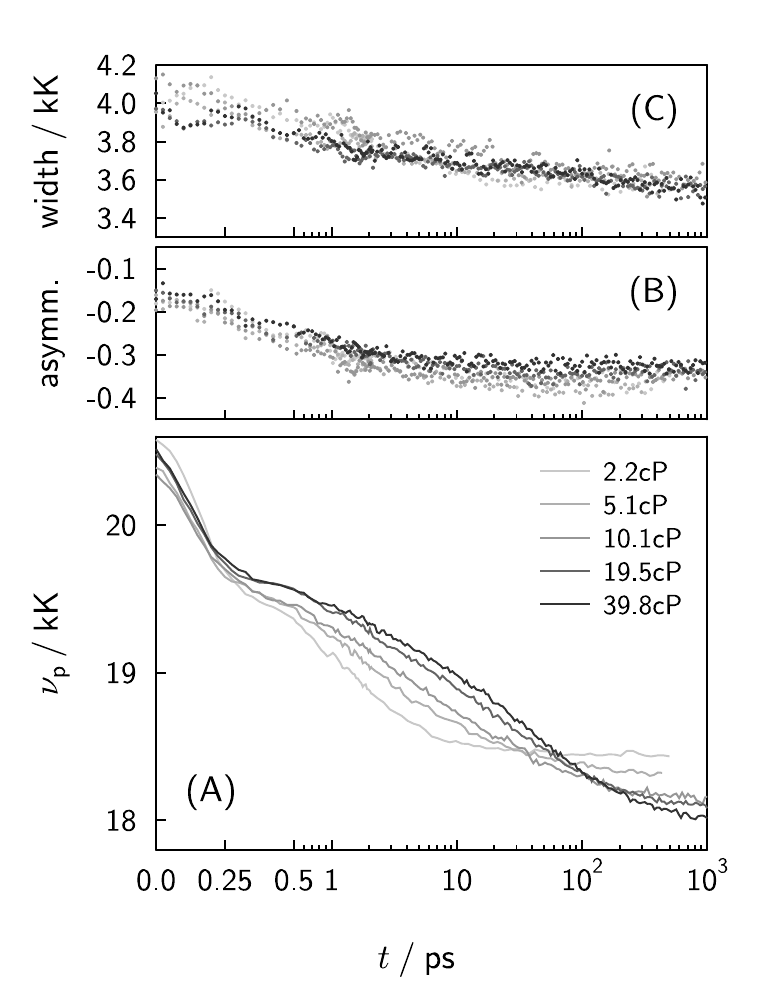}
	\caption{Time-dependence of the (A) maximum and (B) the lineshape asymmetry and (C) linewidth of the fluorescence spectrum of C153 in different binary mixtures of DMSO/GLY at 20$^\circ$C. Note the change from linear to logarithmic time-axis at \unit[0.5]{ps} and that \unit[1]{kK} is equivalent to \unit[1000]{cm$^{-1}$}.}
\label{fig:Ct}
\end{figure}

Further details about the solvation of dyes can be obtained from the dynamic measurement of fluorescence. According to the works of Maroncelli and co-workers, the solvation dynamics of C153 is insensitive to the H-bonding ability of the medium.\cite{horng_JPC_1995} The total dynamic Stokes shift of C153 in various mixtures of DMSO/GLY can be seen in Fig.~\ref{fig:Ct} (see the SI for further expanded experimental results). In all cases, the initial position of the fluorescence maximum, $\nu(0)$, is similar - within the error of the measurement associated to the uncertainties introduced by the chirp compensation and time resolution of the measurement. The overall evolution of the peak shift correlates well with the viscosity of the medium, as can be seen from the average relaxation-times listed in Table~\ref{tab:Ct}. The behaviour within the first \unit[0.5]{ps} is almost identical in all solvents in terms of absolute shift and rate (compare Fig.~\ref{fig:Ct} and Tab.~\ref{tab:Ct}). We decided to analyze $\nu_{\rm p}(t)$ for times longer than \unit[0.5]{ps}, as the dynamics at shorter times can not be described by exponential or Gaussian decays, partly due to the limited time-resolution of the set-up.\footnote{This does not mean that the physics underlying these short times can not be explained by these functions, but rather that without a much shorter excitation pulse available, it is difficult to make proper assessments of the very early dynamics.}\\ 
The shortest of the fitted components are only slightly sensitive to the viscosity (with the shortest, non-fitted, decay component being almost viscosity independent) while the longest vary significantly. The time-dependent spectral lineshapes, on the other hand, do not show any appreciable dependence on solvent composition (cf.\ Fig.~\ref{fig:Ct}). In particular, the lineshape becomes increasingly asymmetric at longer times with the linewidth concomitantly decreasing.\\
Another difference to be considered is the final peak position, $\nu(\infty)$: if solvent relaxation were purely due to dipolar effects (and thus describable by continuum models), the energy of the final state should be the same at all values of $x_{\rm GLY}$. Nonetheless, we find, that the larger the GLY-content the smaller this value, or in other words, the larger is the total solvent relaxation, $\Delta\nu_{\rm total}$ experienced by the probe in the excited state. In view of our findings concerning the KT parameters, this could be attributed to the change in the H-bonding ability. Not having any acidic group, C153 will mostly sense $\alpha$, mainly through the keto group. It is therefore possible that H-bonding with GLY relaxes the excited state of C153 additionally. This is congruent with the findings that the emission maxima of C153 in protic solvents are slightly red-shifted with respect to those in aprotic solvents.\cite{horng_JPC_1995} Ideally, in order to test this hypothesis one should make use of a dye with considerable change of dipole moment upon excitation to the first singlet state, but without any sensitivity to H-bonding. Unfortunately we could not find such a dye, as in order to fulfill the first of the conditions donor and acceptor groups are used that always present some sensitivity to specific solvation.\\

\begin{table*}[!htp]
	\begin{center}
	\begin{threeparttable}[b]
	\caption{Multiexponential fitting parameters to $\nu_{\rm p}(t)$ (eq.\eqref{eq:St}) and some related quantities.}
			\label{tab:Ct}
	\begin{tabular}{c.ccc.cccccc.}
		\toprule
		$x_{\rm GLY}$ & \multicolumn{1}{c}{$\eta$} & \multicolumn{7}{c}{$\nu_{\rm p}(t|t>0.5{\rm ps})$} & $\nu(0)$ & $\Delta \nu_{\rm fast}$ & $\Delta\nu_{\rm total}$&  \multicolumn{1}{c}{$\braket{\tau}_{t>0.5{\rm ps}}$}\\
		 \cmidrule{3-9}
		 &  & $\Delta \nu_1$ & $\tau_1$ & $\Delta \nu_2$ & \multicolumn{1}{c}{$\tau_2$} & $\Delta \nu_3$ & $\tau_3$ &  $\nu(\infty)$  &  &  & & \\
		 &   \multicolumn{1}{c}{(cP)} & (kK) & (ps)  &  (kK) & \multicolumn{1}{c}{(ps)}  &  (kK) & (ps)  & (kK) &   (kK)  & (kK) & (kK) &  \multicolumn{1}{c}{(ps)}  \\ 
		\midrule
		0.00 & 2.2 & 0.66 & 0.9 & 0.52 & 4.5 &  &  & 18.46 & 20.63 & 0.99 & 2.17 & 2.5 \\ 
		0.15 & 5.1 & 0.62 & 1.6 & 0.46 & 11.2 & 0.18 & 122 & 18.30 & 20.40 & 0.84 & 2.11 & 22.4 \\ 
		0.26 & 10.1 & 0.56 & 2.1 & 0.53 & 18.0 & 0.30 & 130 & 18.15 & 20.37 & 0.83 & 2.21 & 35.6 \\ 
		0.36 & 19.5 & 0.51 & 2.1 & 0.64 & 23.9 & 0.39 & 165 & 18.10 & 20.51 & 0.87 & 2.40 & 52.4 \\ 
		0.47 & 39.8 & 0.42 & 2.6 & 0.69 & 30.2 & 0.48 & 203 & 18.02 & 20.50 & 0.90 & 2.49 & 75.4 \\
		\bottomrule
	\end{tabular}
				\begin{tablenotes}
		\item [] $\nu(0)$ is the peak position at time 0. $\Delta \nu_{\rm fast} = \nu(0) - (\nu(\infty) + \sum_i \Delta\nu_i )$, denoting the shift due to ultrafast, not analyzed, components. $\Delta \nu_{\rm total} = \nu(0) - \nu(\infty)$ is the total shift of the peak maximum. $\braket{\tau}_{t>0.5{\rm ps}} = (\sum_i \Delta \nu_i \tau_i)/(\sum_i \Delta\nu_i)$, i.~e.\ the mean lifetime without contributions from the first \unit[0.5]{ps}. \unit[1]{kK} = \unit[1000]{cm$^{-1}$}
		\end{tablenotes}
\end{threeparttable}
	\end{center}
\end{table*}
 
\subsubsection{Rotational Diffusion}
\begin{figure}
\centering
	\includegraphics[scale=1.5]{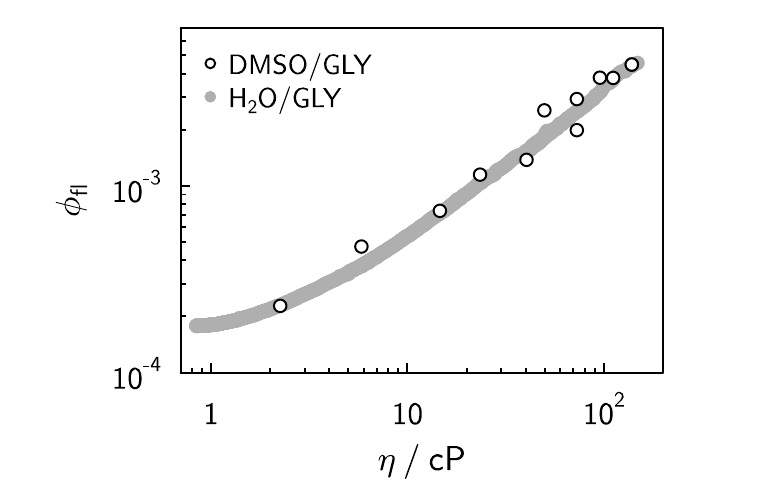}
	\caption{Dependence of the fluorescence quantum yield of Auramine O, $\phi_{\rm fl}$, as a function of solvent viscosity for water/GLY (taken from ref.~\cite{hasegawa_CSA_1996}) and DMSO/GLY mixtures.}
\label{fig:AurO}
\end{figure}

Another way to probe the friction felt by molecular probes in liquid solution is to measure the rotational diffusion of either the entire molecule or of part of it. Auramine O, for example, has flexible moieties capable of undergoing large amplitude motions leading to efficient internal conversion and thus decreasing the fluorescence quantum yield. As the rotating groups are relatively bulky, the solvent viscosity strongly influences the yield of emission. Here, we make use of the systematic study\cite{hasegawa_CSA_1996} of this effect for water-GLY mixtures.\footnote{The last two curves of figure 2 in this article were used. The first one was not correctly printed: Hasegawa, private communication.} The relative fluorescence intensity change of Auramine O with increasing $x_{\rm GLY}$ can be used to check whether the local friction exerted by both solvent mixtures (water/GLY and DMSO/GLY) matches the corresponding macroscopic viscosity measurements. Fig.~\ref{fig:AurO} shows that this is the case.\\

\begin{figure}[!ht]
\centering
	\includegraphics[scale=1.5]{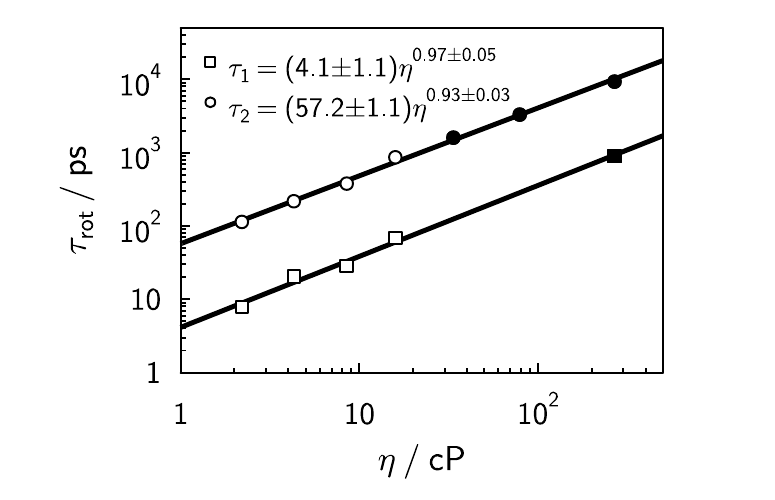}
	\caption{Dependence of the rotational lifetime of C153 as a function of solvent viscosity for the DMSO/GLY mixtures. Open symbols denote measurements with the up-conversion set-up, while filled symbols denote measurements performed with TCSPC. Note, that the short lifetime of the two low viscous samples measured with the TCSPC is not shown, as the limited time-resolution did not allow their proper extraction.}
	\label{fig:tauR}
\end{figure}

A more detailed study of microscopic friction can be performed recording the rotational dynamics of a solute, again C153, monitored via the decay of the fluorescence anisotropy. The starting anisotropy, always larger than 0.36, shows no appreciable dependence on $x_{\rm GLY}$, as expected from the fact that the monitored ${\rm S}_0 \leftarrow {\rm S}_1$ transition does not change with solvent composition. The anisotropy decay in all solvent mixtures exhibits biphasic behavior. Both associated relaxation times increase with the same slope in a double logarithmic plot with the viscosity, despite differing by more than one order of magnitude (cf.\ Fig.~\ref{fig:tauR}, for the experimental decays, refer to the SI).\footnote{The reader is referred to e.~g.\ reference \citenum{qiu_JPCB_2015} for an example with preferential solvation in binary mixtures and the ensuing dynamics.} This indicates that rotational dynamics are rather insensitive to H-bonding, following the same trend as viscosity, as otherwise a discrepancy should be observed in the trends of these times. These present findings are in contrast to those obtained in refs.~\citenum{kaur_JPCB_2014, koley_PCCP_2014}. The possible origin of this discrepancy will be discussed later on. The fact that the anisotropy decays bi-exponentially is connected to the non-sphericity of the probe.\cite{christensen_JPC_1986} Considering a disc-like molecular shape, at least two rotational diffusion coefficients are expected with the rotational times being a combination of them. However, common hydrodynamic models cannot explain this large difference in rotational times observed here. This phenomenon was observed in the past by other groups in single component solvents and several explanations were given.\cite{maroncelli_JCP_1987, horng_JPCA_1997,  sajadi_PCCP_2011, ito_JPCA_2002} A detailed study of these coefficients is beyond the scope of this article as it requires a detailed analysis that would not support further the conclusions already extracted from the data. It is worth noting, that the rotational correlation times in the binary mixtures are congruent with those obtained for C153 in dipolar solvents.\cite{horng_JPCA_1997} The fact that in the former work only monophasic decays in aprotic solvents were observed, can be traced back to the larger experimental noise in these experiments compared to the present data.\\

\subsubsection{Translational Diffusion}
\begin{figure}
\centering
	\includegraphics[scale=1.5]{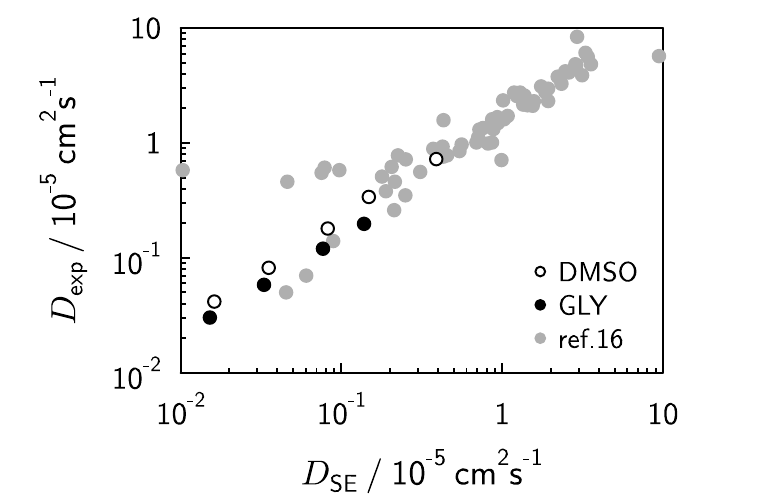}
	\caption{Experimental self-diffusion coefficients of GLY and d6-DMSO in the binary mixtures at 25$^\circ$C versus the Stokes-Einstein value under stick boundary condition, $D_{\rm SE}$. Also shown are the self-diffusion coefficients of a variety of organic solvents (grey full circles) taken from ref.~\citenum{marcus_1998}.}
\label{fig:NMR}
\end{figure}

Finally, another important quantity of special relevance for diffusion influenced reactions is the translational diffusion coefficient, $D$. A relatively straightforward way to test for the trends of this quantity with $x_{\rm GLY}$ is by measuring the self-diffusion coefficients of the solvent components by pulsed field gradient NMR. This technique is not only commonly used for this task,\cite{kaintz_JPCB_2013, araque_JPCB_2015} but also routinely used to assess micro-environments.\cite{carof_JCP_2015, kumarsahu_JPCB_2015, timachova_M_2015, titze_ACIE_2015} Our measurements have been performed both for d6-DMSO and GLY. As can be seen in Fig.~\ref{fig:NMR}, the trends in both cases are similar, scale well with the known self-diffusion coefficients for pure solvents and are directly proportional to the Stokes Einstein value, $D_{\rm SE} = k_{\rm B}T/(6\pi\eta R_{\rm vdw})$, with the van der Waals radius, $R_{\rm vdw}$, given by $R_{\rm vdw} = \sqrt[3]{3/(4\pi) V_{\rm vdw}}$. It can be safely concluded from these measurements that in the time scale probed by NMR there are no deviations from the continuum solvent picture in these mixtures.

\subsection{Comparison to Previous Results}
Several recent works have reported on the presence of micro-heterogeneities in DMSO/GLY mixtures.\cite{chattoraj_JCP_2013, kaur_JPCB_2014, koley_PCCP_2014} In particular, refs.~\citenum{kaur_JPCB_2014} and \citenum{koley_PCCP_2014} report on the solvation dynamics similar to the one we present here, with ref.~\citenum{koley_PCCP_2014} being carried out for 3 different coumarins in the full molar fraction range of GLY. The major differences between the experimental results presented in our work and in ref.~\citenum{koley_PCCP_2014} are concerned with i) temporal and ii) spectral resolution:
\begin{enumerate}[i.]
\item The data in ref.~\citenum{koley_PCCP_2014} were obtained using TCPSC with an IRF of \unit[70]{ps} FWHM, thus limiting the observed spectral relaxation to \unit[200-300]{cm$^{-1}$} for the range of GLY content studied in our work. Compared to the present results, obtained with \unit[170]{fs} time-resolution, this amounts to less than 15\% of the total dynamic Stokes shift, in the best case.
\item The spectral shifts in ref.~\citenum{koley_PCCP_2014} were obtained by applying the spectral reconstruction method\cite{maroncelli_JCP_1987} using 15-18 time-traces (for a spectral range of \unit[6]{kK}, corresponding to a point density of \unitfrac[3]{points}{kK}). The broadband fluorescence experiments presented here, on the other hand, consist of a total of 350 data-points for a spectral range of \unit[10]{kK} (corresponding to a point density of \unitfrac[33]{points}{kK}) with the additional benefit of relying on a robust method for photometric calibration.\cite{zhang_RSI_2011} 
\end{enumerate}
The observed disparities in the results are mostly owed to these significant experimental differences. For example, in addition to an ultrafast decay three relaxation times are here observed - except for pure DMSO - while in ref.~\citenum{koley_PCCP_2014} only two are listed. The relaxation times presented now range from \unit[1-3]{ps} (absent in ref.~\citenum{koley_PCCP_2014}), over \unit[5-30]{ps} upon increasing viscosity (also absent in ref.~\citenum{koley_PCCP_2014}), up to the longest relaxation time ranging from \unit[122-203]{ps}. This latter component loosely matches the fastest resolved component of \unit[88-420]{ps} ref.~\citenum{koley_PCCP_2014}. The longest component of \unit[257-4485]{ps} in ref.~\citenum{koley_PCCP_2014}, which corresponds to amplitudes of the dynamic Stokes shift (in the case of C153) of \unit[7-130]{cm$^{-1}$}, is absent in the present observations. In any case, even if these latter shifts are measurable, which is quite difficult, given the low point density of the experiments in ref.~\citenum{koley_PCCP_2014}, they would merely correspond to 0.3-6\% of the total dynamic Stokes shift.\footnote{Comparing the steady-state emission maxima, $\nu_{\infty}$ (\unit[18.46-18.02]{kK}), with their steady-state analoga, $\nu_{\rm ss}$ (\unit[18.50-18.30]{kK}), presented here, it is found that the former are consistently slightly red-shifted with respect to the latter, as expected. Therefore it is very unlikely that any part of the relaxation is missed in the present dynamic measurements.}\\
Our findings are also partially at odds with similar data of Ghosh and co-workers on anisotropy decays of coumarins in DMSO-GLY mixtures.\cite{kaur_JPCB_2014, koley_PCCP_2014} Using TCSPC with an IRF of \unit[70]{ps} biexponential anisotropy decays were found. On the one hand, one of the rotational lifetimes remains almost constant and equal to the rotational time observed in pure DMSO. On the other hand, the second and longer component scales with viscosity and roughly agrees with our results. Here we want to emphasize that in the present case the time resolution is by almost 3 orders of magnitude higher than for the experiments reported in refs.~\citenum{kaur_JPCB_2014, koley_PCCP_2014}. The present experiments thus allow resolving the viscosity dependence of the short component (see Figure~\ref{fig:tauR}), which is - at least at low and intermediate GLY content - significantly below the time resolution of the other reports. In addition, the paper does not clarify how the rotational lifetimes have been obtained, as for anisotropy decays with rotational lifetimes in the range of the IRF dedicated fitting routines are required in order to recover correct lifetimes.\cite{barkley_JCP_1981, soutar_M_1996} This is the reason why in the present work the fit to the short components of the TCPSC at the lower viscosities measured with this technique, was not performed (cf. Figure~\ref{fig:tauR}). It is thus possible, that the observed discrepancies in the analysis and interpretation of the sub-100~ps rotational lifetimes are a consequence of the limited time-resolution in the experiments of refs.~\citenum{kaur_JPCB_2014, koley_PCCP_2014}. Besides, the long rotational lifetimes, as found by Koley et al.\ at large GLY content, are to be treated with care, as soon as they significantly exceed the fluorescence lifetime of the probe molecule.\\
We are not commenting the results of the other two coumarins measured in ref.~\citenum{koley_PCCP_2014} as they seem to be more prone to specific solvation, what may be mixed-up with preferential solvation. We want to take the opportunity to emphasize that one should not mix the concepts of preferential solvation or dielectric enrichment,\cite{suppan_JCSFT1_1987, krolicki_JPCA_2002} micro-heterogeneity (which is probe-independent),\cite{grant_JPCB_2005} and specific solvation,\cite{catalan_LA_1996, catalan_LA_1997, marcus_1998} as seems to be the case in the commented papers.\\
In reference \citenum{chattoraj_JCP_2013} it was concluded that the mixtures of DMSO and GLY at viscosities of 10 and \unit[33]{cP} show heterogenities. The experiments consisted of fluorescence correlation spectroscopy (FCS) and burst integrated fluorescence lifetime (BIFL) measurements obtained from fluorescence microscopy on four coumarin dyes (C151, C152, C152A and C153).\footnote{We should mention, that from the information in ref.~\citenum{chattoraj_JCP_2013} it remains ambiguous whether C153 or C480 was used, as the chemical structure corresponds to C153, while the attributed name is C480.} From both experimental techniques bimodal distributions for the translational diffusion coefficient (from FCS) and for the fluorescence lifetime (from BIFL) of the coumarins were extracted. However, the distributions obtained for different coumarins as well as comparing the two techniques (for the same coumarin) differed. These differences were attributed to the fact that FCS monitors long range movements while BIFL probes only a small environment, sensed during the excited state of the probes. Following this line of reasoning of having significantly different environments, the anisotropy decays - even though measured on the bulk sample - should also be sensitive to the heterogeneities. In fact, fluorescence anisotropy decays have been extensively used in the past to discuss heterogeneous media and preferential solvation.\cite{krolicki_JPCA_2002, shirota_JCP_2000, kim_JPCA_2004, grant_JPCB_2005, pal_TCA_2013} More precisely, if two different environments are felt by the probe in the excited state, as proposed in ref.~\citenum{chattoraj_JCP_2013} for the DMSO/GLY mixtures, two \emph{long} rotational times should be observed, one belonging to a DMSO-rich environment and the second to a bulk like one, with the former being invariant and the latter changing with solvent composition. However, the present experiments show two relaxation times, which distinctly and systematically change with solvent viscosity, and no trace of another relaxation time corresponding to pure DMSO, constant over the range of molar fractions, is observed.\\
The diffusion coefficients obtained from averaging the results of FCS in ref.~\citenum{chattoraj_JCP_2013} amount roughly to two to four times the expected value from the Stokes-Einstein equation depending on the coumarin. Such deviations should be noticeable in the bulk measurements as those reported here from the NMR experiments or from the anisotropy decays, but are not observable in the presented experiments (see Figs.~\ref{fig:tauR} and \ref{fig:NMR}).\\
Another test for micro-environments can be made by rationalizing simply the total population decay of the excited state of the probes. In ref.~\citenum{chattoraj_JCP_2013}, BIFL experiments yielded, especially for C152, two Gaussian lifetime distributions centered at around 1 and approx.~\unit[4]{ns} respectively. This should translate into multiexponential fluorescence decays\footnote{In the limit of narrow distributions the observed fluorescence decays simplify to sums of as many exponentials as maxima in the distribution, while in the case of broad gaussian distributions the fluorescence decays should follow functions described in ref.\,\citenum{berberan_CP_2005, berberan_CP_2005a, berberan_JL_2007}.} as detected by bulk measurements of TCSPC.\cite{berberan_CP_2005, berberan_CP_2005a, berberan_JL_2007, boens_PPS_2014} However, this is not the case neither for C153, nor for C152 in none of the mixtures explored (see SI). We have in fact tried to account for a 2$^{\rm nd}$ long lifetime, by fixing the lifetimes to the values observed in ref.~\citenum{chattoraj_JCP_2013} and leaving only their amplitudes free (see SI). We found that the resulting amplitudes were negligibly small and the additional lifetime did not improve the resulting reduced $\chi^2$. There is a change in the lifetime of emission of these molecules, which most likely is due to differences on the non-radiative decays as the H-bonding properties of the solvent changes.\cite{rosspeintner_JACS_2012} We have, nevertheless, no explanation for the observed multimodal distributions observed in ref.~\citenum{chattoraj_JCP_2013}, though it is quite intriguing that the largest secondary contributions observed are for C152 and C152A, both with lifetimes around \unit[1]{ns}, while for the other coumarins, with a lifetime of \unit[5]{ns} the secondary distributions are almost negligibly small. At the same time, the additional lifetime for C152 and C152A, in ref.~\citenum{chattoraj_JCP_2013}, amount to \unit[4-5]{ns}, which is not observable for these compounds in polar solvents or in the presently discussed mixtures.\cite{nad_JPCA_2003}

\section{Conclusions}
The combination of DMSO and GLY represents an interesting binary solvent mixture for studying changes associated to viscosity in both intra- and intermolecular chemical reactions, as it is the observed property exhibiting the most pronounced change upon changing the molar fraction. The mixture's dielectric properties remain almost constant, especially in terms of the Onsager function. This renders the mixture especially interesting with respect to electron transfer reactions, or, in general, processes controlled by the dielectric relaxation properties of the medium. However two cautions need to be taken: H-bonding properties of the mixture also vary greatly, and the dielectric relaxation is rather complex. The former means that it is wiser to use reactants with no or little sensitivity to H-bonding, i.~e.\ small $a$ in the KT terms. The latter implies, that whenever the process is still controlled at long times by dielectric relaxation, the composition of the mixture will change these times as well as viscosity. On the other hand, processes, which are fast enough to sense only the first, and most relevant, part of the relaxation of the dielectric, as is commonly thought to be the case for electron transfer reactions,\cite{weaver_JPC_1990, weaver_CR_1992} will experience solvent relaxation times that are almost invariant with the composition.\\
According to recent molecular dynamics simulations additional caution has to be taken when the solutes are smaller than the solvent molecules, such as methane.\cite{araque_JCP_2015} As in the present measurements this case has not been addressed, with the investigated solute molecules being significantly larger, we did not detect these deviations. This is in line with the conclusions reached by Araque et al.\ in reference \citenum{araque_JCP_2015}.\\
From the former conclusions it is now well understood why these mixtures have been useful in the study of the diffusion-influence in photo-induced electron transfer in the past.\cite{neufeld_JCP_2002, angulo_JPCA_2003, gladkikh_PCCP_2003, gladkikh_JPCA_2007, rosspeintner_CEJ_2007, angulo_CEJ_2010, koch_JACS_2012, rosspeintner_JACS_2012} In these papers, a reaction-diffusion model\cite{burshtein_ACP_2004} in combination with Marcus theory\cite{marcus_JCP_1956} for electron transfer was able to explain the experimental observations \emph{quantitatively}. Finally, in clear contrast to previous reports, we have found no experimental evidence for micro-heterogeneities in all the measurements presented here.

 \section*{Acknowledgement}
G.~A.\ acknowledges financial support from the Narodowe Centrum Nauki within the ``Harmonia 3" program, grant number 2012/06/M/ST4/00037. M.~G.\ acknowledges financial support by the DFG through SFB 1078. A.~R.\ and E.~V.\ thank Nikolaus Ernsting for help with the FLUPS set-up and acknowledge financial support from the SNF (No.\ 200020-147098) and the University of Geneva.
\appendix

\clearpage
\section{Appendix - Samples}

\begin{figure}[h!]
\centering
	\includegraphics{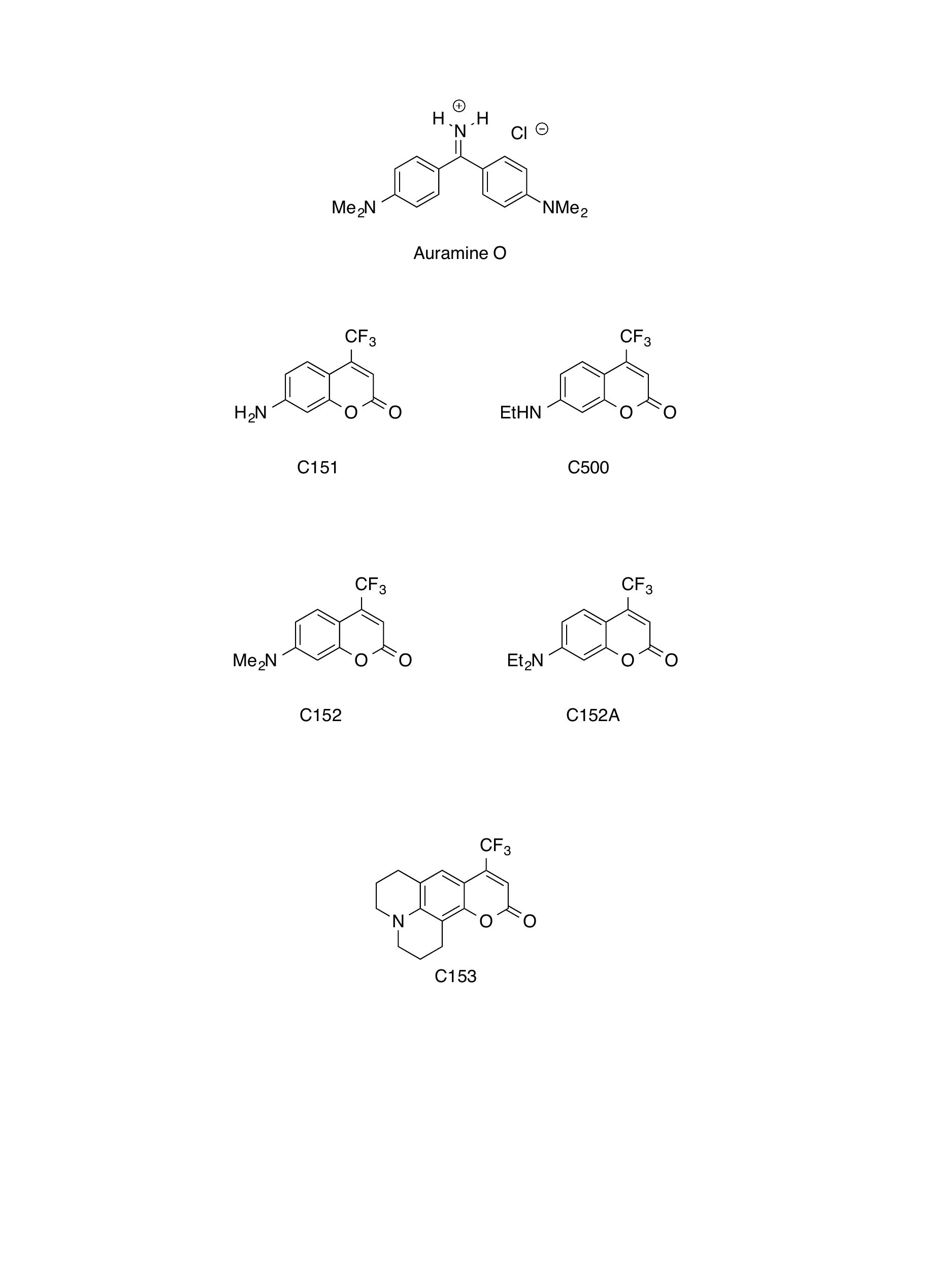}
\caption[Structures of the used probes]{Structures of the used probes. Note, that C152 has only been used for experiments in the SI.}
\end{figure}

\clearpage
\section{Appendix - Absorption Spectra}

\begin{figure}[h!]
    \subfloat[Absorption spectra]{%
      \includegraphics[width=0.49\textwidth]{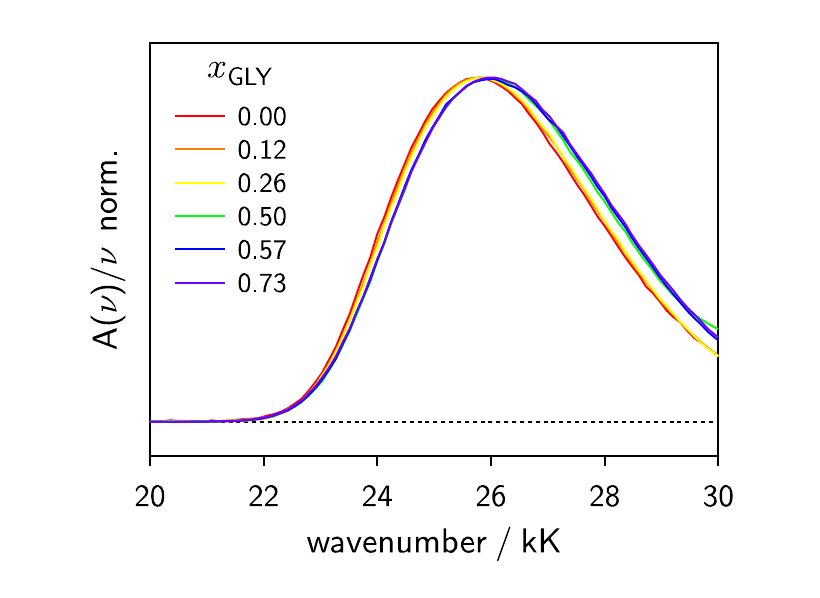}
    }
    \hfill
    \subfloat[Shifted absorption spectra]{%
      \includegraphics[width=0.49\textwidth]{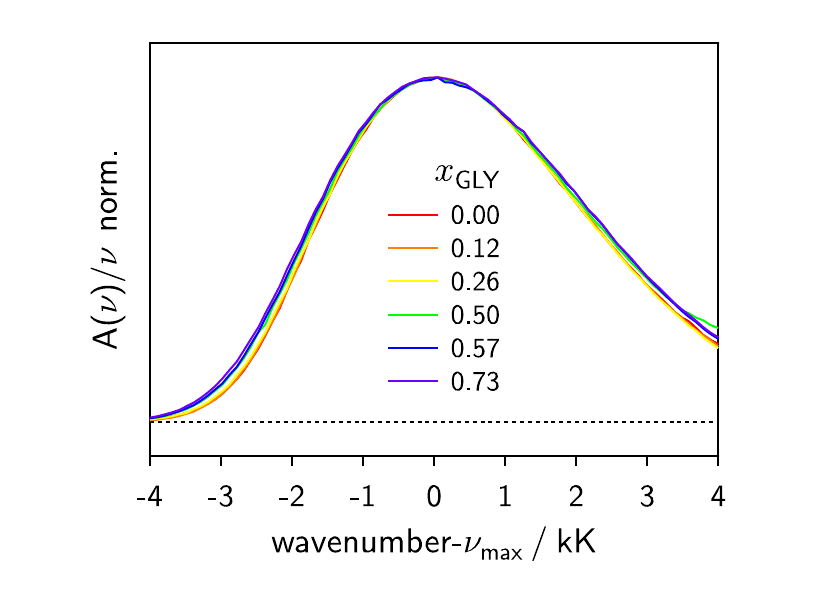}
    }
	\caption[Absorption spectra of C151]{Transition dipole moment representation of the absorption spectra of C151 in mixtures of DMSO/GLY. The original spectra (left) and the spectra, shifted by the position of their maximum, $\nu_{\rm max}$, (right), are shown.}
\label{fig:C151_abs}
\end{figure}

\begin{figure}[h!]
    \subfloat[Absorption spectra \label{subfig-2:dummy}]{%
      \includegraphics[width=0.49\textwidth]{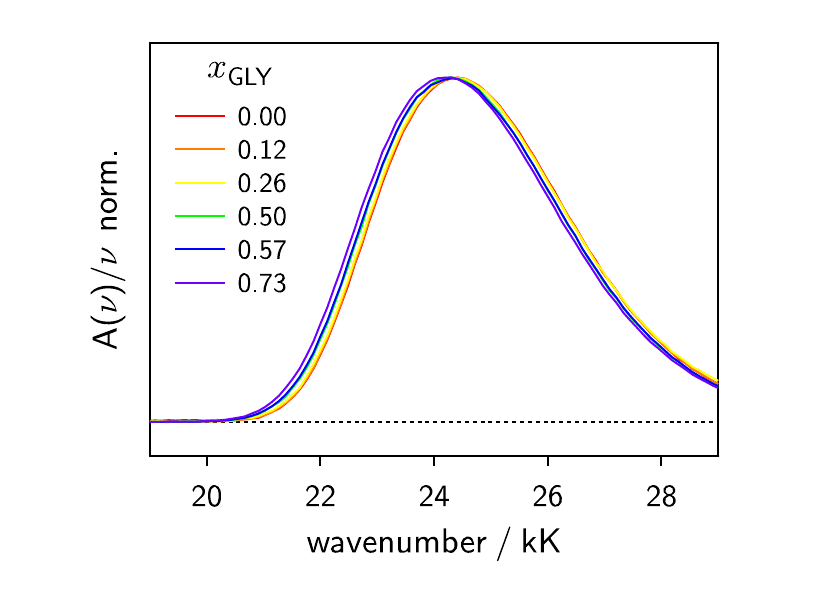}
    }
    \hfill
    \subfloat[Shifted absorption spectra \label{subfig-1:dummy}]{%
      \includegraphics[width=0.49\textwidth]{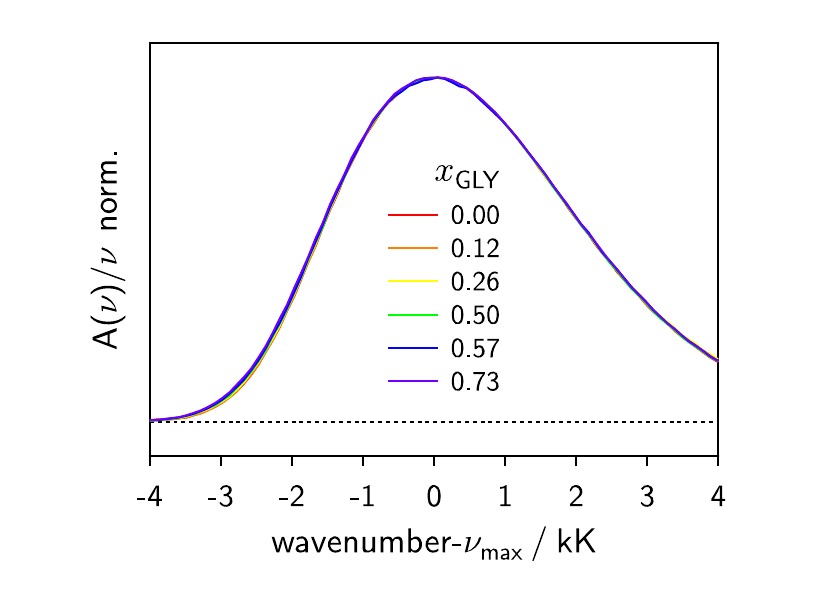}
    }
	\caption[Absorption spectra of C152A]{Transition dipole moment representation of the absorption spectra of C152A in mixtures of DMSO/GLY. The original spectra (left) and the spectra, shifted by the position of their maximum, $\nu_{\rm max}$, (right), are shown.}
\label{fig:C152A_abs}
\end{figure}

\begin{figure}[h!]
    \subfloat[Absorption spectra \label{subfig-2:dummy}]{%
      \includegraphics[width=0.49\textwidth]{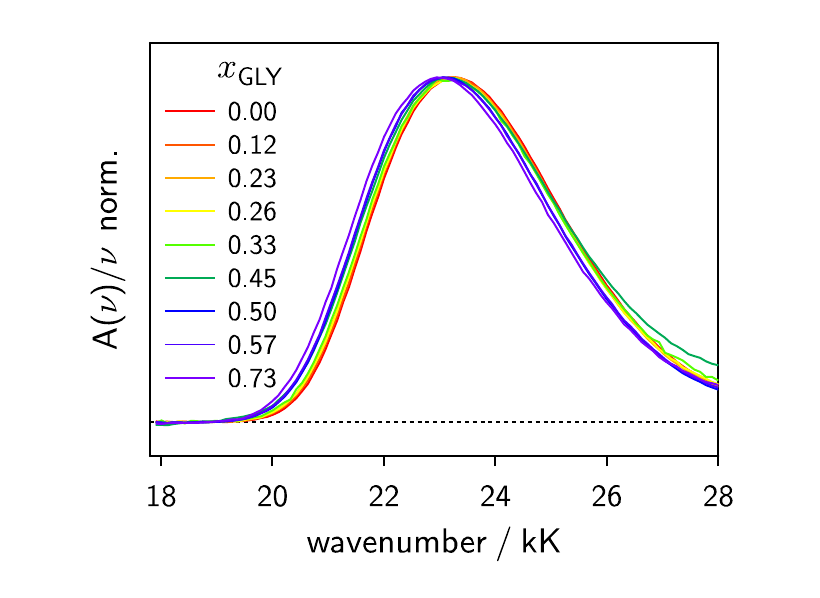}
    }
    \hfill
    \subfloat[Shifted absorption spectra \label{subfig-1:dummy}]{%
      \includegraphics[width=0.49\textwidth]{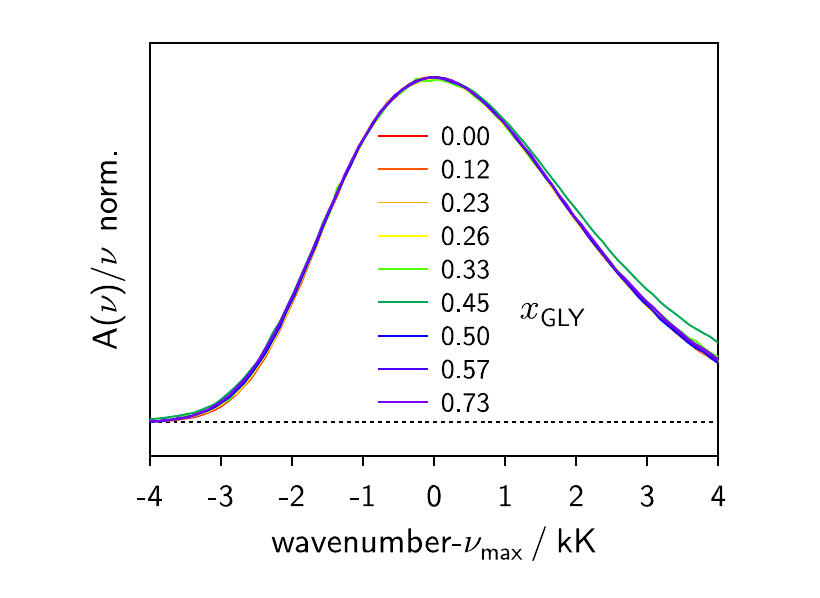}
    }
	\caption[Absorption spectra of C153]{Transition dipole moment representation of the absorption spectra of C153 in mixtures of DMSO/GLY. The original spectra (left) and the spectra, shifted by the position of their maximum, $\nu_{\rm max}$, (right), are shown.}
\label{fig:C500_abs}
\end{figure}

\begin{figure}[h!]
    \subfloat[Absorption spectra \label{subfig-2:dummy}]{%
      \includegraphics[width=0.49\textwidth]{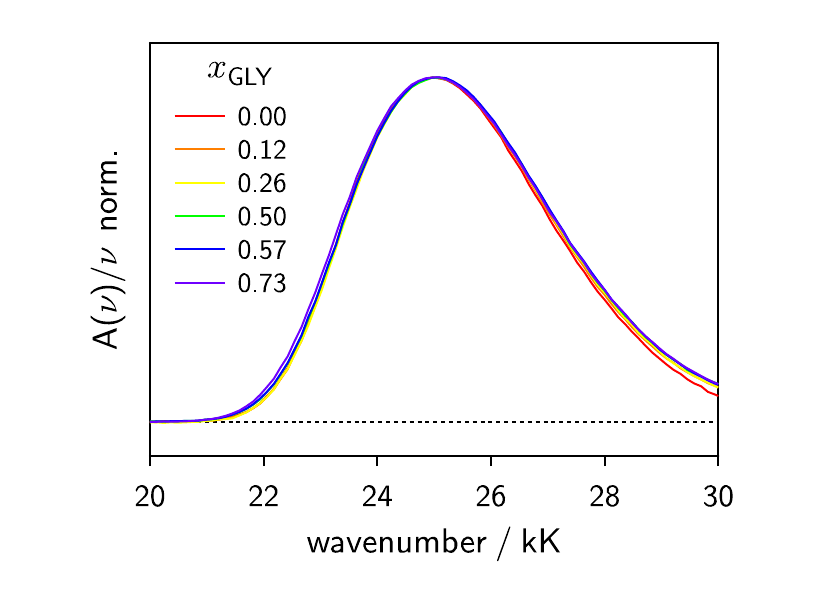}
    }
    \hfill
    \subfloat[Shifted absorption spectra \label{subfig-1:dummy}]{%
      \includegraphics[width=0.49\textwidth]{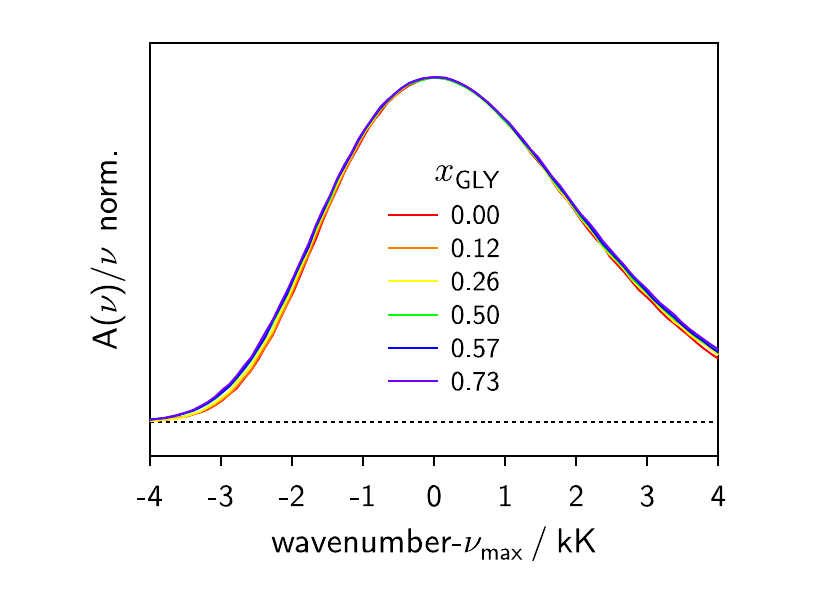}
    }
	\caption[Absorption spectra of C500]{Transition dipole moment representation of the absorption spectra of C500 in mixtures of DMSO/GLY. The original spectra (left) and the spectra, shifted by the position of their maximum, $\nu_{\rm max}$, (right), are shown.}
\label{fig:C500_abs}
\end{figure}

\clearpage
\section{Appendix - Kamlet-Taft Analysis}

\begin{figure}[!h]
   \centering
      \subfloat[Coumarin 151]{\includegraphics{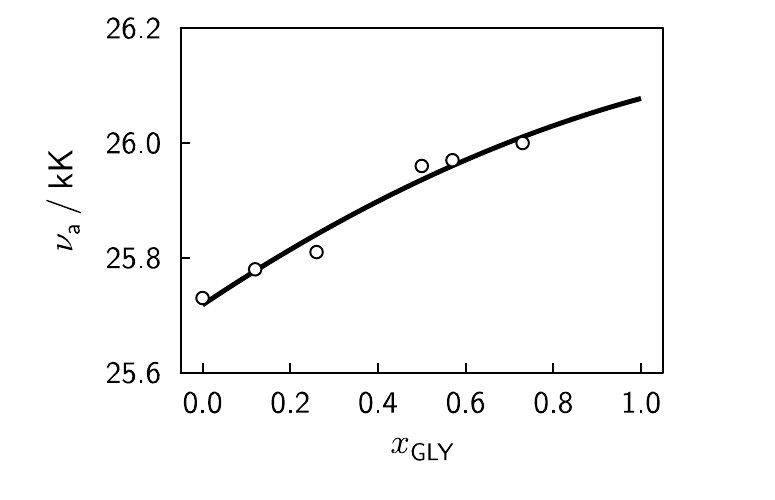}}
      \subfloat[Coumarin 152A]{\includegraphics{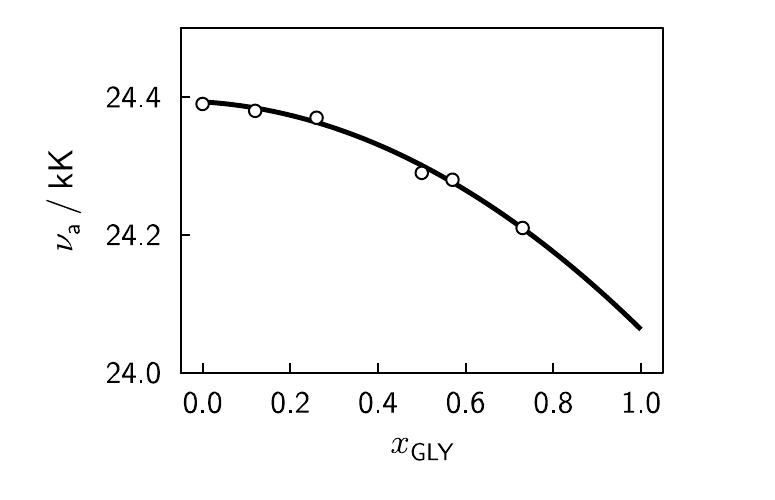}}\\
      \subfloat[Coumarin 153]{\includegraphics{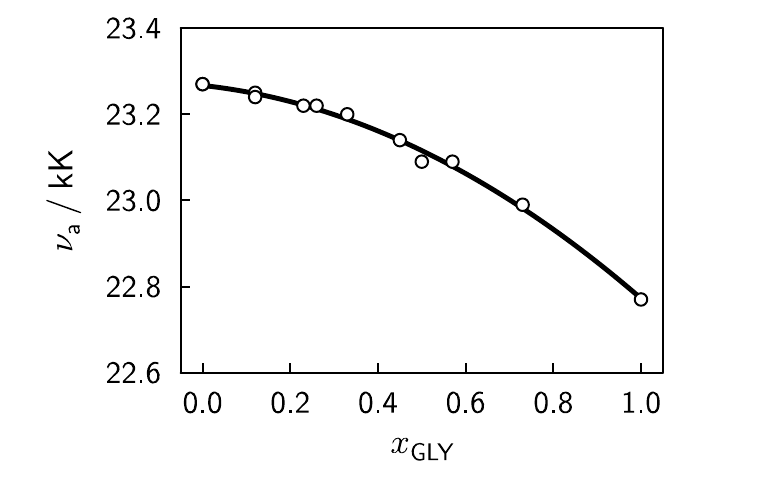}}
	\subfloat[Coumarin 500]{\includegraphics{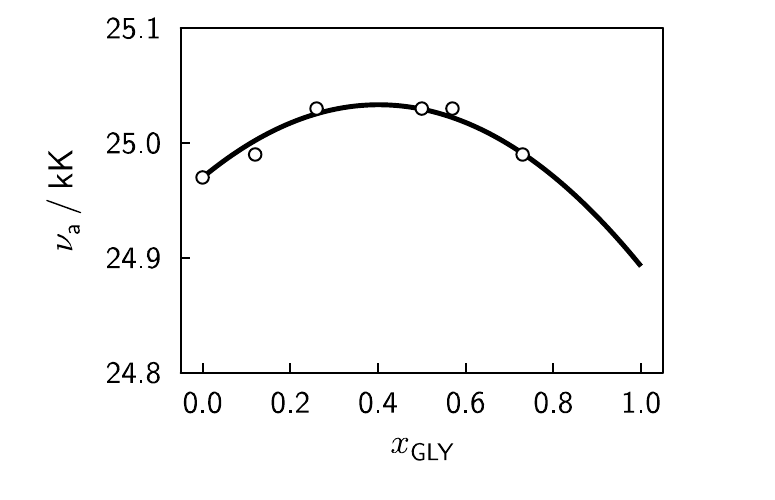}}
   \caption[Peak position of absorption vs.\ solvent composition]{Dependence of the position of the low energy absorption maximum on the molar fraction of glycerol. Solid lines denote polynomial fits of 2$^{\rm nd}$-order with the parameters from Table\,\ref{tab:KT_pars}.}
\end{figure}

\begin{table}[!h]
	\begin{center}
	\setlength{\tabcolsep}{0.5cm}
	\caption[Fitting parameters for absorption peak position vs.\ solvent composition]{Fitting parameters for the dependence of the low energy absorption maximum on the molar fraction of glycerol using $\nu_{\rm a} = a x_{\rm GLY}^2 + b x_{\rm GLY} + c$.}
	\begin{tabular}{l;;;}
			\toprule
			 & \multicolumn{1}{c}{$a$} & \multicolumn{1}{c}{$b$} & \multicolumn{1}{c}{$c$} \\
			 & \multicolumn{1}{c}{(kK)} & \multicolumn{1}{c}{(kK)} & \multicolumn{1}{c}{(kK)} \\
			 \midrule
		C151 & -0.152 & 0.512 & 25.72 \\ 
		C152A & -0.292 & -0.038 & 24.39 \\ 
		C153 & -0.384 & -0.111 & 23.27 \\ 
		C500 & -0.393 & 0.315 & 24.97\\
		\bottomrule
		\end{tabular}
		\label{tab:KT_pars}
	\end{center}
\end{table}

\clearpage
\section{Appendix - Fluorescence Lifetimes}

\begin{figure}[h]
    \subfloat[C152  at  2.2 cP \label{subfig-2:dummy}]{%
      \includegraphics[width=0.49\textwidth]{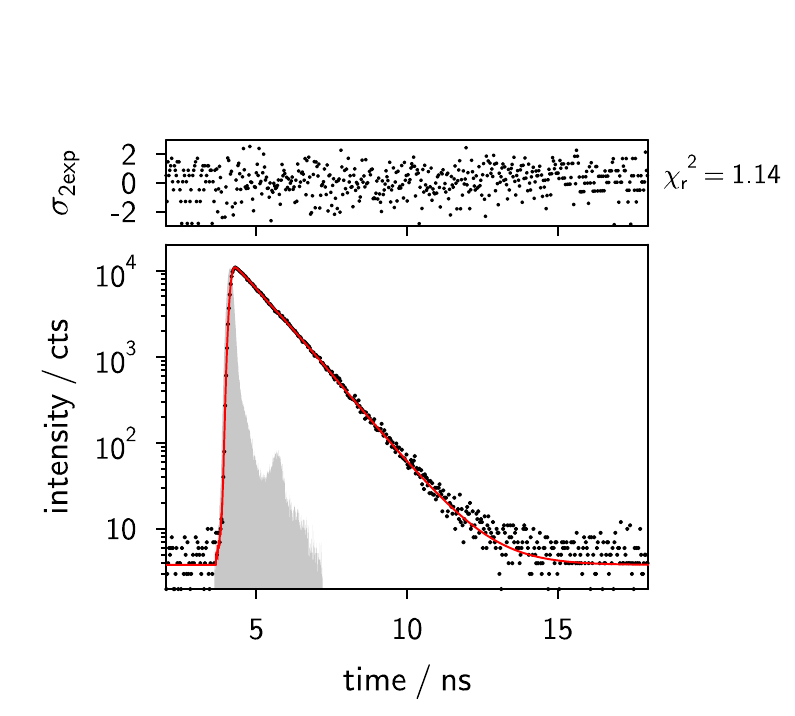}
    }
    \hfill
    \subfloat[C152 at 10 cP \label{subfig-1:dummy}]{%
      \includegraphics[width=0.49\textwidth]{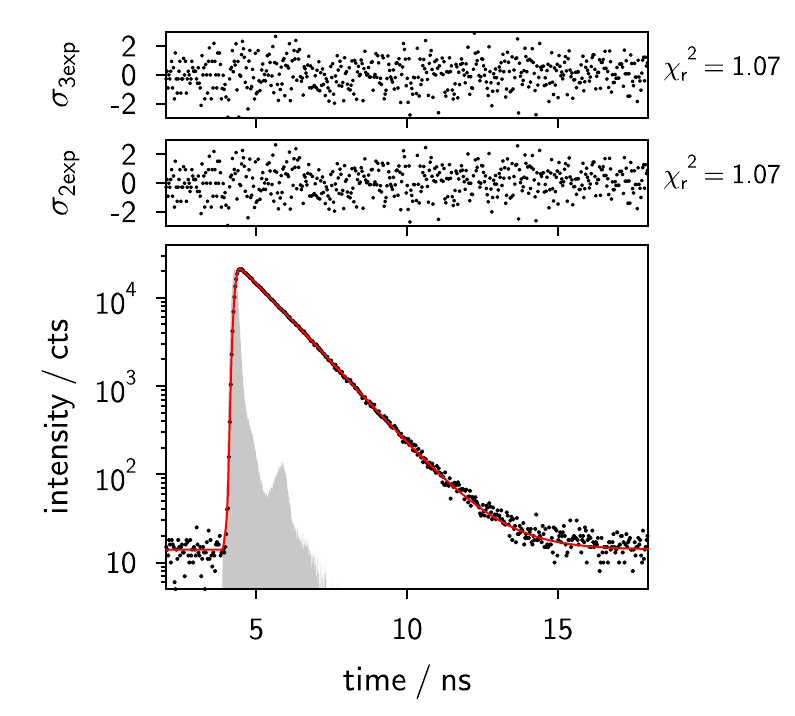}
    }\\
    \subfloat[C152  at 50cP \label{subfig-2:dummy}]{%
      \includegraphics[width=0.49\textwidth]{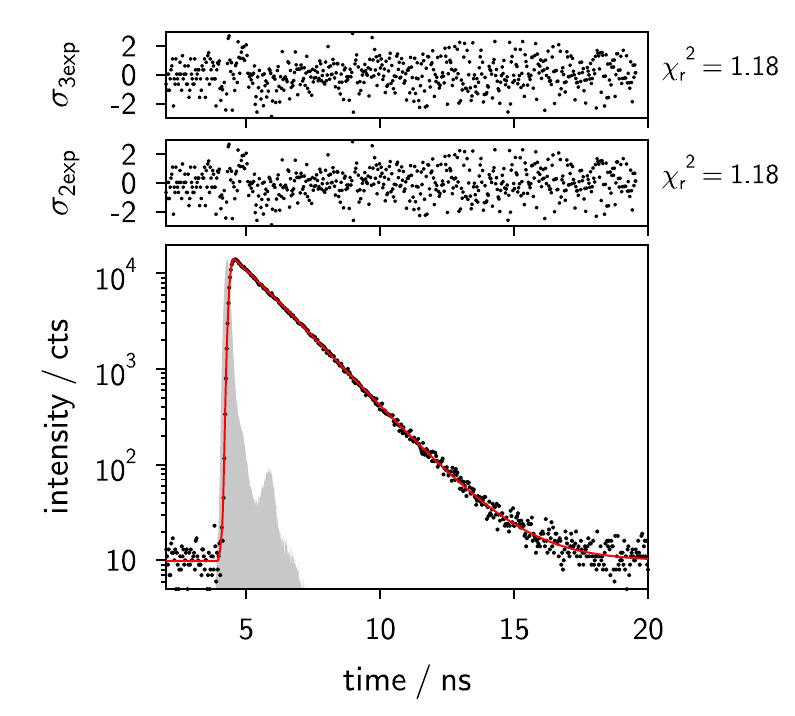}
    }
	\caption[Sub-nanosecond fluorescence decays of C152 (at 2.2, 10, 50cP)]{Fluorescence decays of C152 in mixtures of DMSO/GLY with varying viscosity. The data were thinned out by a factor of 5 for better visualization. The weighted residuals correspond to the best fits using the parameters given in Table\,\ref{tab:tcspc}. The grey area denotes the instrument response function.}
\label{fig:tcspc_C152}
\end{figure}

\begin{figure}[h]
    \subfloat[C153  at  2.2 cP \label{subfig-2:dummy}]{%
      \includegraphics[width=0.49\textwidth]{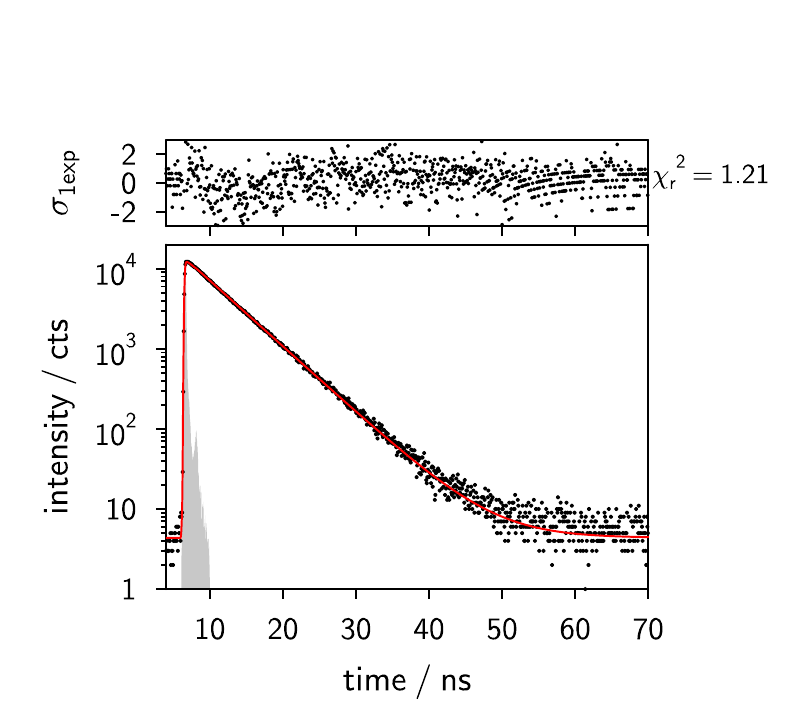}
    }
    \hfill
    \subfloat[C153 at 10 cP \label{subfig-1:dummy}]{%
      \includegraphics[width=0.49\textwidth]{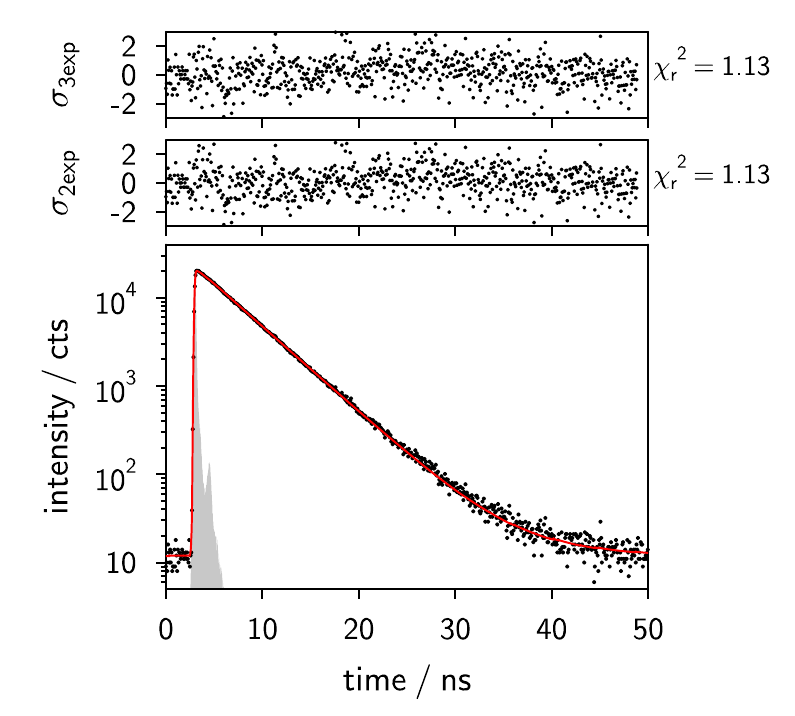}
    }\\
    \subfloat[C152  at 40cP \label{subfig-2:dummy}]{%
      \includegraphics[width=0.49\textwidth]{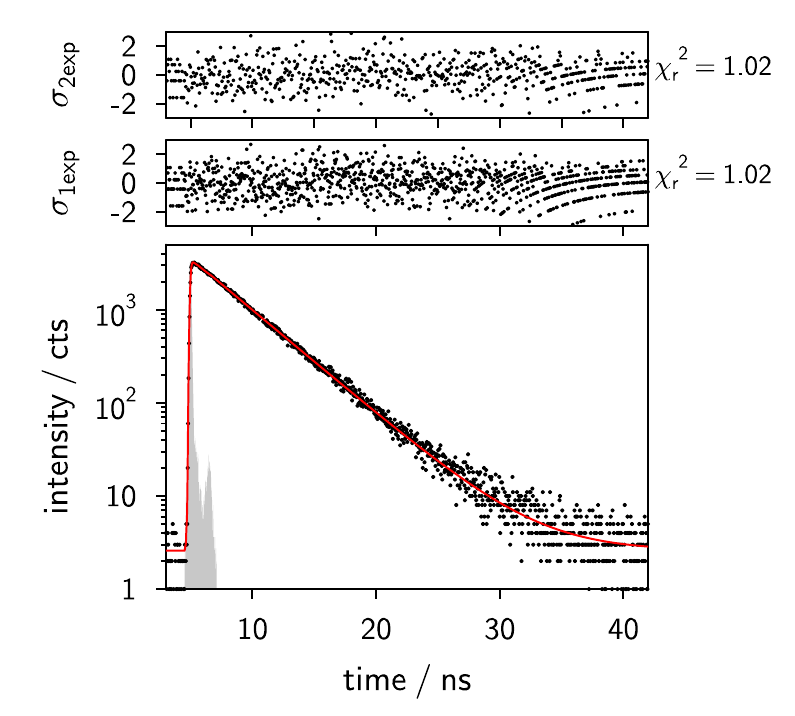}
    }
    \hfill

	\caption[Sub-nanosecond fluorescence decays of C153 (at 2.2, 10, 40cP)]{Fluorescence decays of C153 in mixtures of DMSO/GLY with varying viscosity. The data were thinned out by a factor of 3 for better visualization. The weighted residuals correspond to the best fits using the parameters given in Table\,\ref{tab:tcspc}. The grey area denotes the instrument response function.}
\label{fig:tcspc_C153}
\end{figure}

In order to further test for possible micro-heterogeneities, as e.\,g.\ reported in ref.~33, we opted for complementary measurements. In particular, the multimodal distributions obtained from burst-integrated fluorescence lifetime (BIFL) experiments in ref.~33, should also show up in bulk measurements of the fluorescence lifetime of the corresponding fluorophores as multiphasic fluorescence decays. Table\,\ref{tab:tcspc} and Figures\,\ref{fig:tcspc_C152} and \ref{fig:tcspc_C153} summarize our findings for two example-fluorophores (C153, as it is the main probe in our study and C152, as the multimodal behaviour has been reported to be very prominent) in three representative solvents (mixtures). Depending on the chosen emission wavelength (using narrowband interference filters) a single exponent or two exponents are necessary to properly describe (as judged from the reduced $\chi^2$) the fluorescence decays. The additional short lifetime (in addition to the nanosecond S$_1$ depopulation time), is necessary to account for features arising from solvation dynamics. However, it should be noticed, that both, the amplitude and lifetime ($< 100$\,ps) of the short component are prone to large errors, given the low time-resolution in our experiments (FWHM of the IRF of approx.\ 200\,ps).

\begin{table}[!h]
	\begin{center}
	\setlength{\tabcolsep}{0.5cm}
	\caption[Sub-nanosecond fluorescence fitting parameters for C152 and C153]{Fitting parameters for fluorescence lifetimes of C152 and C153 presented in Figures\,\ref{fig:tcspc_C152} and \ref{fig:tcspc_C153}. Note, that the parameters from the biexponential fit, and the third lifetime have been fixed in the 3-exponential fit.}
	\begin{tabular}{;cccccc}
			\toprule
		\multicolumn{1}{c}{$\eta$}	&  $\tau_1$ & $A_2/A_1$ & $\tau_2$ & $A_3/A_1$ & $\tau_3$ & $\chi_{\rm r}^2$ \\ 
		\multicolumn{1}{c}{(cP)} &	 (ns) &  & (ns) &  & (ns) &  \\
			 \midrule
			&  \multicolumn{6}{c}{C152}\\\cmidrule{2-7}			 
		2.2 & (0.01) &  (0.10) & 1.05 & & & 1.14\\			 
		10 & 0.06 &  3.8 & 1.17 & & & 1.07\\
			&	  0.06 &  3.8 & 1.17 & $9\cdot 10^{-4}$ & 4.0 & 1.07\\ 
		50 & 0.06 &  1.56 & 1.49 & & & 1.18\\
			&	 0.06 &  1.56 & 1.49 & $6\cdot 10^{-6}$ & 4.0 & 1.18\\[1.5ex]
						&  \multicolumn{6}{c}{C153}\\\cmidrule{2-7}
		2.2 &  5.27 &  &  & & & 1.21\\			 
		10 & 0.04 &  2.42 & 4.47 & & & 1.13\\
			& 0.04 &  2.42 & 4.47 & $9\cdot 10^{-3}$ & 2.0 & 1.13\\ 
		40 & 3.90 &   &  & & & 1.02\\		 
		& 3.90 &   &  & $3\cdot 10^{-3}$  & 1.5 & 1.02\\
		\bottomrule
		\end{tabular}
		\label{tab:tcspc}
	\end{center}
\end{table}

\clearpage
\section{Appendix - Fluorescence Anisotropy}

\begin{figure}[!h]
\centering
\includegraphics[scale=1.5]{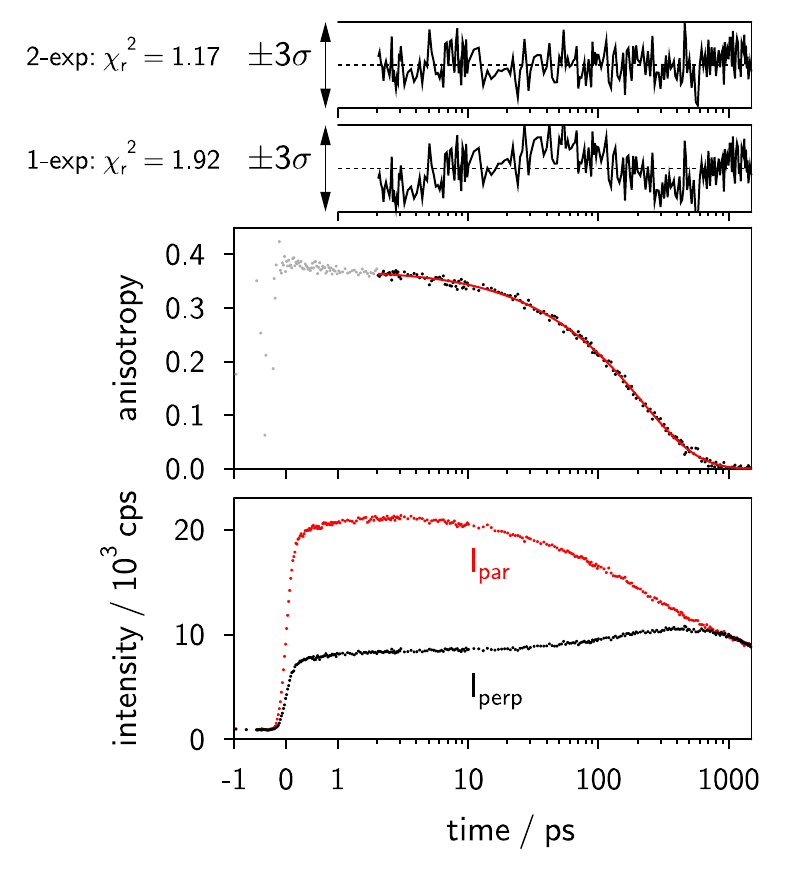}
\caption[Femtosecond fluorescence decays and ensuing anisotropy of C153 at 4.3cP]{Fluorescence up-conversion traces of C153 in a DMSO/GLY mixture with $x_{\rm GLY} = 0.12$, upon excitation at \unit[405]{nm} and detection at \unit[540]{nm} under parallel (I$_{\rm par}$) and perpendicular (I$_{\rm perp}$) excitation polarization. The resulting anisotropy is also shown, together with a fit to the data for times longer than \unit[2]{ps} and the ensuing weighted residuals for a single exponential or bi-exponential fit.}
\label{}
\end{figure}

\begin{figure}[!h]
\centering
\includegraphics[scale=1.5]{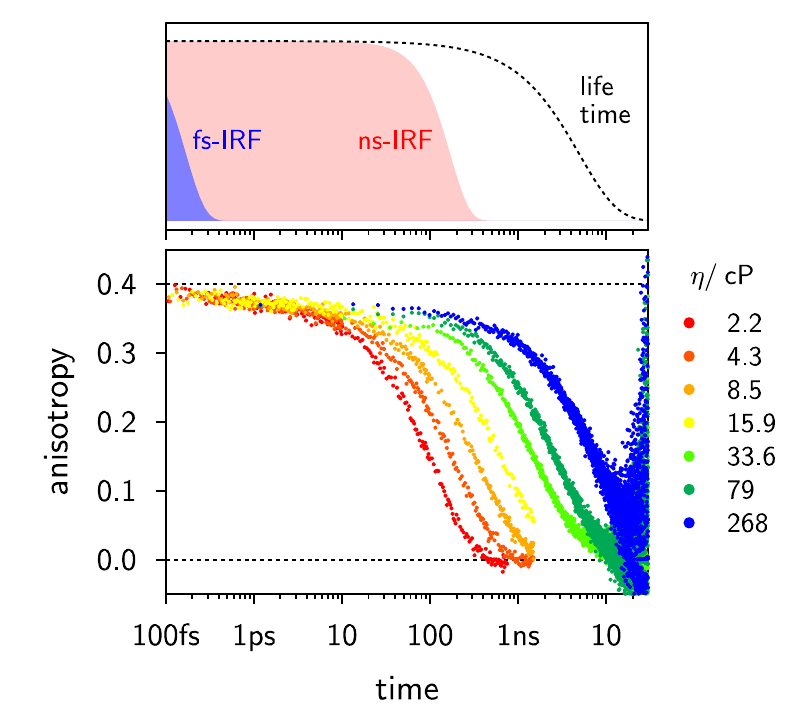}
\caption[Anisotropy decays of C153 in mixtures of DMSO/GLY]{Comparison of fs- and ns-anisotropy decays of C153 in mixtures of DMSO/GLY of different viscosities. The upper panel indicates the instrument response functions of the 2 set-ups used and the approximate fluorescence lifetime. The three high viscosity samples were measured using ns-TCSPC.}
\label{}
\end{figure}

\begin{table}[!h]
	\begin{center}
	\setlength{\tabcolsep}{0.5cm}
	\caption[Fitting parameters for anisotropy decays]{Fitting parameters for fluorescence anisotropy decays of C153. The sum of the amplitudes, $a_i$ equals 1. Note, that the short component for the TCSPC for the two low viscous mixtures is only tentative, as these times are in the range of the IRF and can thus not be properly extracted using the simple fitting approach.}
	\begin{tabular}{;cc;cc}
			\toprule
		\multicolumn{1}{c}{$\eta$}	& $r(0)$ & $a_1$ & \multicolumn{1}{c}{$\tau_1$} & $\tau_2$ & $\chi_{\rm r}^2$ \\ 
		\multicolumn{1}{c}{(cP)} &	 & & \multicolumn{1}{c}{(ps)} &  (ps) &  \\
			 \midrule
			&  \multicolumn{5}{c}{FOG}\\\cmidrule{2-6}			 
		2.2 &  0.39 & 0.07 & 8 & 110 & 1.04\\	
		4.3 &  0.37 & 0.08 & 20 & 220 & 1.07 \\	
		8.5 &  0.38 & 0.09 & 28 & 380 & 1.31 \\
		15.9 & 0.38 & 0.12 & 79 & 860 & 1.03\\[1ex]							
			&  \multicolumn{5}{c}{TCSPC}\\\cmidrule{2-6}
		33.6 & 0.33 & 0.22 & \multicolumn{1}{c}{(200)}  & 1600 & 1.12 \\	
		79 & 0.37 & 0.24 & \multicolumn{1}{c}{(400)} & 3300 & 1.36\\	
		268 & 0.41 & 0.20 & 910 & 9300 & 1.21\\				 
		\bottomrule
		\end{tabular}
		\label{tab:tcspc}
	\end{center}
\end{table}

\clearpage
\section{Appendix - Comparison of Broadband Fluorescence Data}

\begin{figure}[!h]
   \centering
      \includegraphics[scale=1.3]{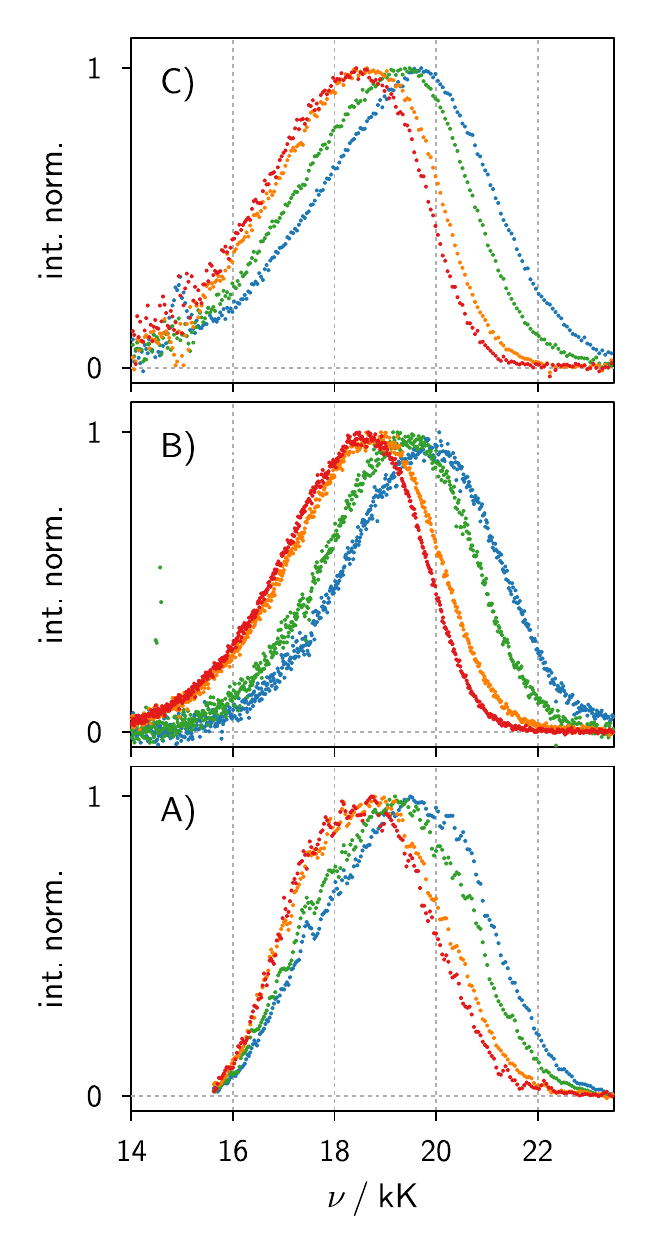}
   \caption[]{Selected broadband fluorescence spectra of C153 in DMSO at 0.3 (blue), 0.8 (green), 5 (orange) and 100\,ps (red) on the broadband fluorescence set-ups in A) Warsaw, B) Berlin and C) Geneva. Note, that no photometric correction has been applied to the data in A) and B) and that the 2 short and 2 long times in B) have been recorded with different crystal thicknesses (leading to a small spectral displacement by approx.\ \unit[0.2]{kK} due to different photometric correction curves). The data in C) have been photometrically corrected using a set of fluorophores, the corrected spectra of which have been recorded on a FluoroMax-4, which itself has been calibrated using a set of secondary emissive standards.}
\end{figure}

\begin{figure}[!h]
   \centering
   \subfloat[Absolute peak position]{\includegraphics[scale=1.3]{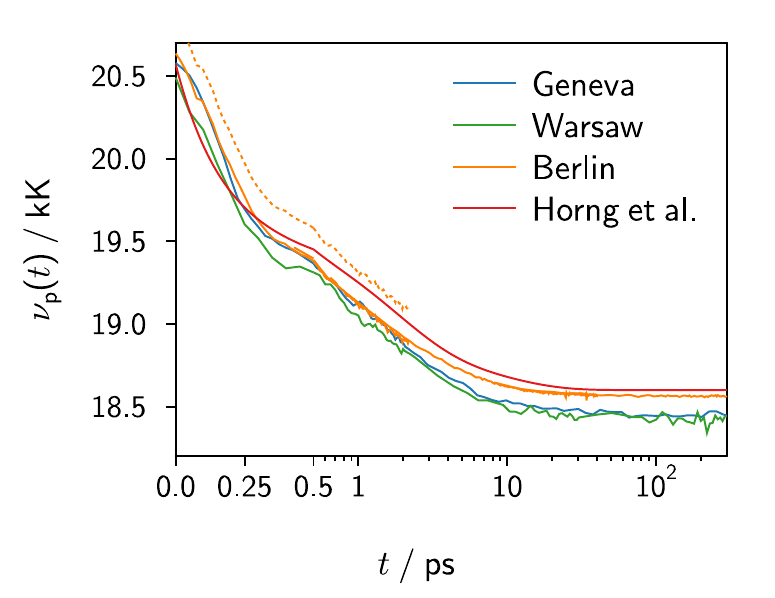}}\\
   \subfloat[Peak shift]{\includegraphics[scale=1.3]{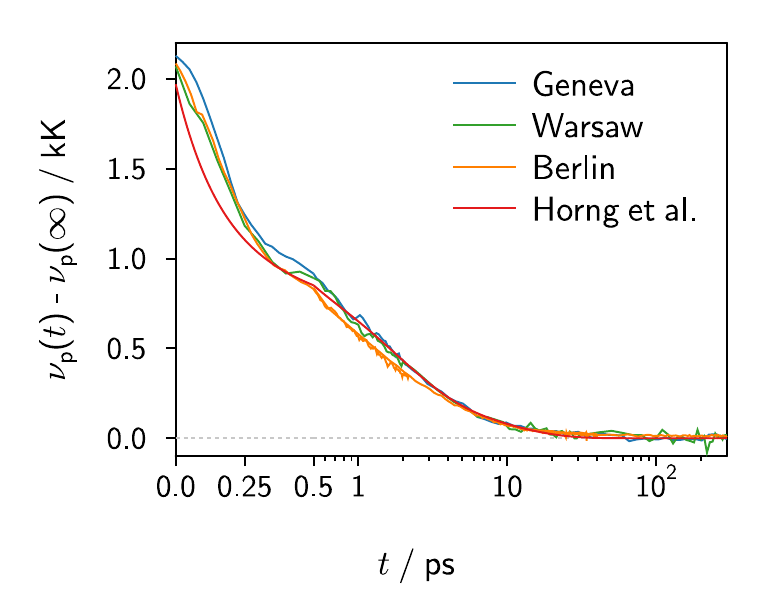}}
   \caption[]{a) Absolute peak position and b) peak shift (i.\,e.\ $\nu_{\rm p}(t) - \nu_{\rm p}(\infty)$) of C153 in DMSO as measured on the 3 broadband fluorescence set-ups. The use of two different crystal thicknesses in the Berlin data introduces a vertical displacement of\unit[0.2]{kK} between the two data-sets (dashed line vs.\ full line). The overlap region (between \unit[0.2-2]{ps}) has been used to match the two data sets, by merely displacing one of them by \unit[0.2]{kK}. In addition the reconstructed curve from reference 4 of the main manuscript (Horng et al.) has been added.}
\end{figure}

\bibliographystyle{achemso}
\bibliography{total}

 \end{document}